\documentclass[11pt,a4paper]{article}

\usepackage[truedimen,margin=34mm]{geometry} 

\usepackage{mathrsfs}
\usepackage{amssymb}
\usepackage{amsmath}
\usepackage{ascmac}
\usepackage{amsthm}
\usepackage[pdftex]{graphicx}
\usepackage[pdftex]{color}
\usepackage{color}
\usepackage{booktabs}
\usepackage{setspace}
\usepackage{natbib}
\usepackage{url}

\usepackage{titlesec}
\titleformat*{\section}{\large\bfseries}
\titleformat*{\subsection}{\it}

\title{{\bf Predicting Infection of COVID-19 in Japan: State Space Modeling Approach}}
\date{}

\begin{document}

\maketitle
\doublespacing

\vspace{-1.5cm}
\begin{center}
Genya Kobayashi$^1$, Shonosuke Sugasawa$^2$, Hiromasa Tamae$^3$ and Takayuki Ozu$^3$

\end{center}

\noindent
$^{1}$Graduate School of Social Science, Chiba University,\\
$^{2}$Center for Spatial Information Science, The University of Tokyo\\
$^{3}$Nospare Inc.
\vspace{5mm}

\begin{center}
{\bf \large Abstract}
\end{center}
The number of confirmed cases of the coronavirus disease (COVID-19) in Japan has been increasing day by day and has had a serious impact on the society especially after the declaration of the state of emergency on April 7, 2020. 
This study analyzes the real time data from March 1 to April 22, 2020 by adopting a sophisticated statistical modeling tool based on the state space model combined with the well-known susceptible-exposed-infected (SIR) model. 
The model estimation and forecasting are conducted using the Bayesian methodology. 
The present study provides the parameter estimates of the unknown parameters that critically determine the epidemic process derived from the SIR model and prediction of the future transition of the infectious proportion including the size and timing of the epidemic peak with the prediction intervals that naturally accounts for the uncertainty. 
The prediction results under various scenarios reveals that the temporary reduction in the infection rate until the planned lifting of the state on May 6 will only delay the epidemic peak slightly.
In order to minimize the spread of the epidemic, it is strongly suggested that an intervention is carried out for an extended period of time and that the government and individuals make a long term effort to reduce the infection rate even after the lifting.


\newpage
%%%%%%%%%%%%%%%%%%%%%%%%%%%%%%%%%%%%%%%%%%
\section{Introduction}
\begin{figure}[htb!]
\centering
\includegraphics[width=8cm]{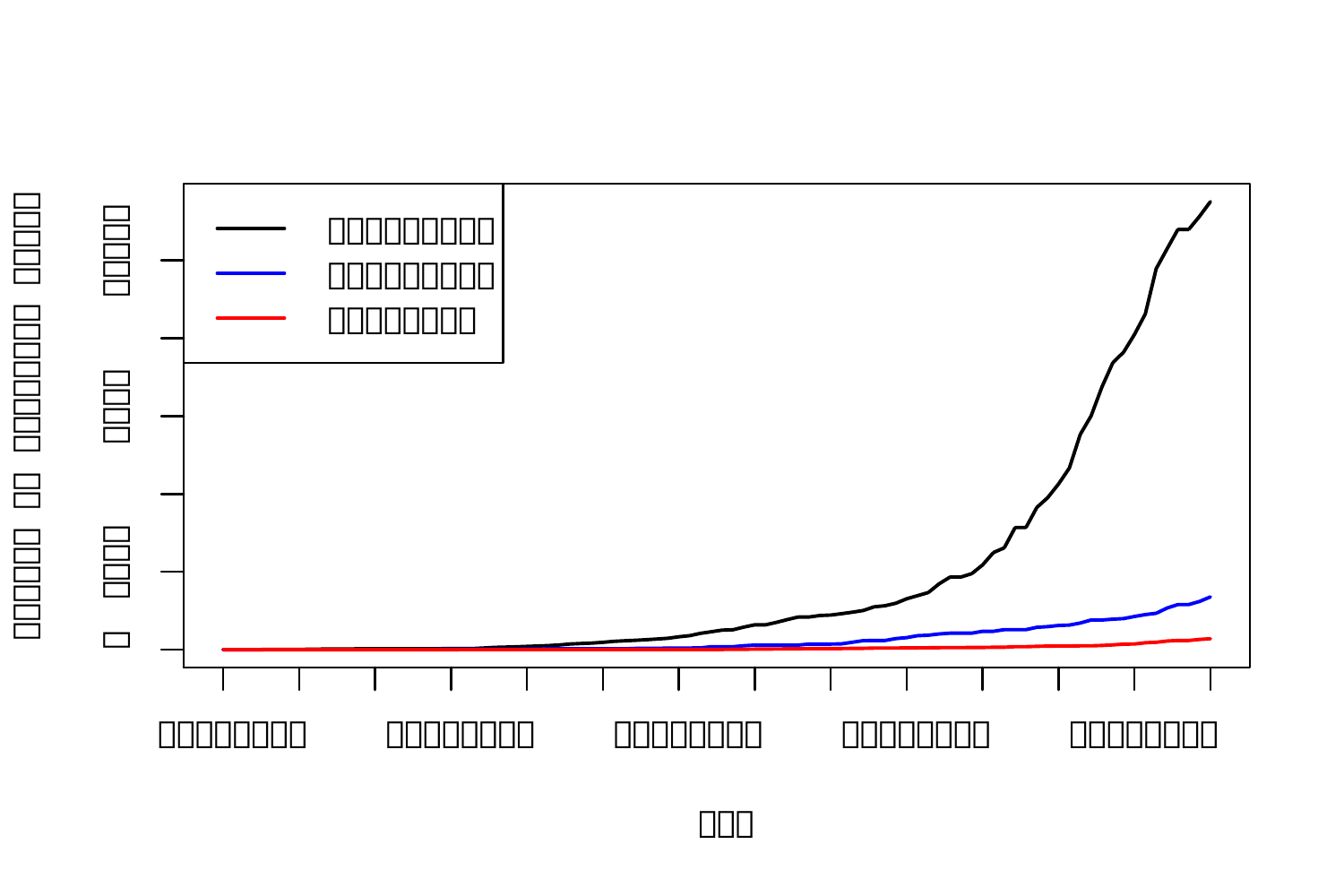}
\caption{The cumulative numbers of confirmed, recovered and deceased persons in Japan. \label{fig:number}}
\end{figure}

Since the first case of the coronavirus disease 2019 (COVID-19) in Japan was confirmed on January 15, 2020, the number of confirmed cases has been increasing day by day. 
Although the Japanese government declared a state of emergency on April 7, it does not have a legal force to regulate individual activities and remains at only requesting the avoidance of outings. 
Consequently, the number of cases has been still increasing as shown in Figure \ref{fig:number}. 
Needless to say, Japanese economy has been seriously shocked and the public interest mainly lies on how the number of infected persons transits in the future and when the outbreak will converge.
Although there already exists a rapidly increasing number of statistical analyses of the epidemic, the statistical evidence focusing on the situations in Japan is still limited except for \cite{Kuniya2020, Karako2020, Mizumoto2020}. 
Therefore, the purpose of this study is to provide a statistical evidence regarding the future transition of the infectious proportion in Japan, including the intensity and timing of the epidemic peak, based on the real-time data on the cumulative number of confirmed, recovered and deceased persons, shown in Figure \ref{fig:number}.

We consider the famous susceptible-infected-recovered (SIR) model \cite{KK1927} for modeling the epidemic process as widely adopted in the existing literature on COVID-19.
However, this deterministic model is not necessarily sufficient to explain the variability of the transition since the observed number is subject to nonignorable randomness. 
To handle such randomness in the data, we employ the state spate models combined with the SIR model (SS-SIR model) developed by \cite{Dukic2012, Osthus2017}. 
The model was originally proposed for statistical modeling of the seasonal trend of influenza. 
The advantages of the SS-SIR model are mainly three points; (1) the unknown parameters in the SIR model can be estimated with little knowledge about the true values by adequately using the data information; (2)  future prediction of a variety of quantities such as the number of infections or the epidemic peak as well as uncertainty quantification of the prediction can be carried out easily; (3) whether the real-time data follows the assumed SIR model or not can be assessed through the parameter estimate. 
These advantages are quite essential because (1)information required for modeling the epidemic trend of a new virus is scarce, (2) it is important to compute not only point prediction but also interval prediction to understand the possible worst and best scenarios of future transition, and (3) understanding if the real-time data actually follows the SIR model is critical for the reliability of future simulations based on the SIR model.

\vspace{-0.5cm}

%%%%%%%%%%%%%%%%%%%%%%%%%%%%%%%%%%%%%%%%%%
\section{Methods}

\subsection{Data}
We use the numbers of confirmed, recovered and deceased persons collected on an open source platform  (\url{https://www.kaggle.com/sudalairajkumar/ novel-corona-virus-2019-dataset}).
Although the original data starts from January 22, the numbers before the end of February are treated collectively.
This is because the confirmed numbers in this period are relatively small and using the data after March 2020 would be useful to reliably predict the future numbers of infectious persons after May 2020. 
Hence, the period of the data used in our analysis consists of $T=53$ days from March 1 to April 22. 
We use the difference between the cumulative numbers of confirmed persons and recovered plus deceased persons, denoted by $Z(t)$ for $t=1\ldots,T$, which can be interpreted as the number of confirmed persons being infectious.
It is further assumed that only $p \ (0<p\leq 1)$ fraction of infectious individuals can be identified by diagnosis, which is called {\it identification rate} hereafter.
Then we define $Y(t)$ as $Z(t)=Np\times Y(t)$ where $N=1.265\times 10^8$ is the population of Japan, thereby $Y(t)$ is the proportion of the infectious population at time $t$.
Regarding the specific values of $p$, we follow the discussion in \cite{Kuniya2020}.
Since \cite{Hokkaido} reported that 77 persons were confirmed among the possible 940 infected population, the $95\%$ confidence interval of $p$ is $(0.059, 0.105)$.
Based on this argument, the results under the following three scenarios $p=0.05$, $0.1$ and $0.2$ are compared.

\subsection{Statistical model}
Here the model proposed by \cite{Osthus2017} is described. 
Let $S(t)$, $I(t)$ and $R(t)$ denote the proportions of individuals being susceptible, infected and recovered population at the time $t$, respectively, satisfying $S(t)+I(t)+R(t)=1$. 
The SIR model describes the epidemic over time via the nonlinear ordinary differential equations (ODE) given by 
\begin{equation}\label{SIR}
S'(t)=-\beta S(t)I(t), \ \ \ \ \ I'(t)=\beta S(t)I(t) -\gamma I(t), \ \ \ \ \ R'(t)=\gamma I(t),
\end{equation}
where the unknown infection rate $\beta>0$ and removal rate $\gamma>0$ control the transition from one compartment to the next and jointly determine the epidemic process. 
Let $\theta(t)=(S(t), I(t), R(t))$ define the three-dimensional vector of the unobserved true proportion at the time $t$. 
To allow randomness in the evolution of $\theta(t)$, the following model is considered:
\begin{equation}\label{state}
\theta(t)|\theta(t-1) \sim {\rm Dir}(\kappa f(\theta(t-1);\beta, \gamma)),  \ \ \ t=1,\ldots,T, 
\end{equation}
where Dir$(\cdot)$ denotes the Dirichlet distribution, $f(\theta(t-1);\beta, \gamma)$ is the solution of the deterministic SIR model (\ref{SIR}) starting the ODE at $\theta(t-1)$ and $\kappa>0$ is the unknown parameter controlling the randomness in the evolution. 
In the above model, the conditional expectation of $\theta(t)$ given the previous state $\theta(t-1)$ is $f(\theta(t-1);\beta, \gamma)$, so the distribution of $\theta(t)$ is centered around the deterministic model (\ref{SIR}).
It is noted that the conditional variance of $\theta_t$ decreases as $\kappa$ increases, thus the validity of the assumption of the deterministic model (\ref{SIR}) can be verified through the estimate of $\kappa$.

Let $Y(t)$ be the observed value of $I(t)$. 
Since $Y(t)$ is not necessarily equal to the true $I(t)$, $Y(t)$ is observed based on the following probabilistic model:
\begin{equation}\label{ob}
Y(t)|I(t)\sim {\rm Beta}(\lambda I(t), \lambda(1-I(t))),  \ \ \ t=1,\ldots,T, 
\end{equation}
where Beta$(\cdot, \cdot)$ denotes the Beta distribution and $\lambda>0$ is an unknown parameter having a similar role to $\kappa$ in (\ref{state}).
The statistical model for $Y(t)$ with the combination of (\ref{state}) and (\ref{ob}) is seen as a {\it state space model}.

The unknown parameters in the model are the two parameters $\beta$ and $\gamma$ in the SIR model and two scale parameters $\kappa$ and $\lambda$ that control the randomness in the two equations (\ref{state}) and (\ref{ob}).
The estimation of these parameters and future prediction is conducted within the Bayesian framework in which we assign prior distributions for these parameters and compute the posterior distribution via the {\it Bayes rule}. 
Due to the complexity of the model, the analytical derivation of the posterior distribution is not feasible. 
Instead, we rely on the simulation-based method known as {\it Markov Chain Monte Carlo} (MCMC) algorithm \citep{gamerman} to generate random numbers from the posterior distribution. 
Then the parameter estimates are calculated and future prediction is carried out based on the output of the MCMC algorithm.
Regarding the prior distributions, we assign slightly non-informative priors to reflect the uncertainty about the new epidemic and let the data tell the truth adequately.
The details of the settings of the prior distributions and algorithm are provided in Supplementary Material.

%%%%%%%%%%%%%%%%%%%%%%%%%%%%%%%%%%%%%%%%%%
\section{Results}
\subsection{Prediction of epidemic peak}
\begin{figure}[htb!]
\centering
\includegraphics[width=4.5cm]{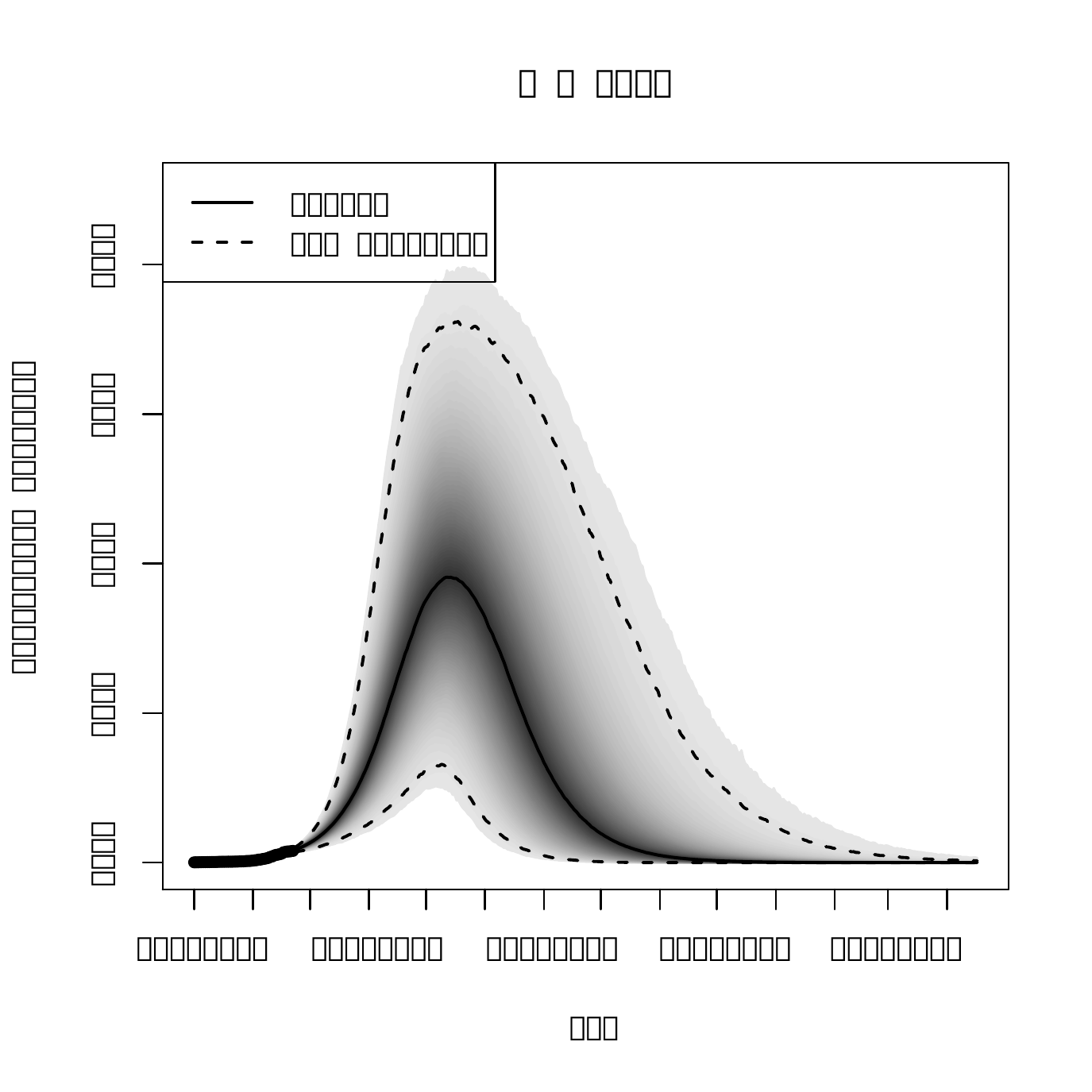}
\includegraphics[width=4.5cm]{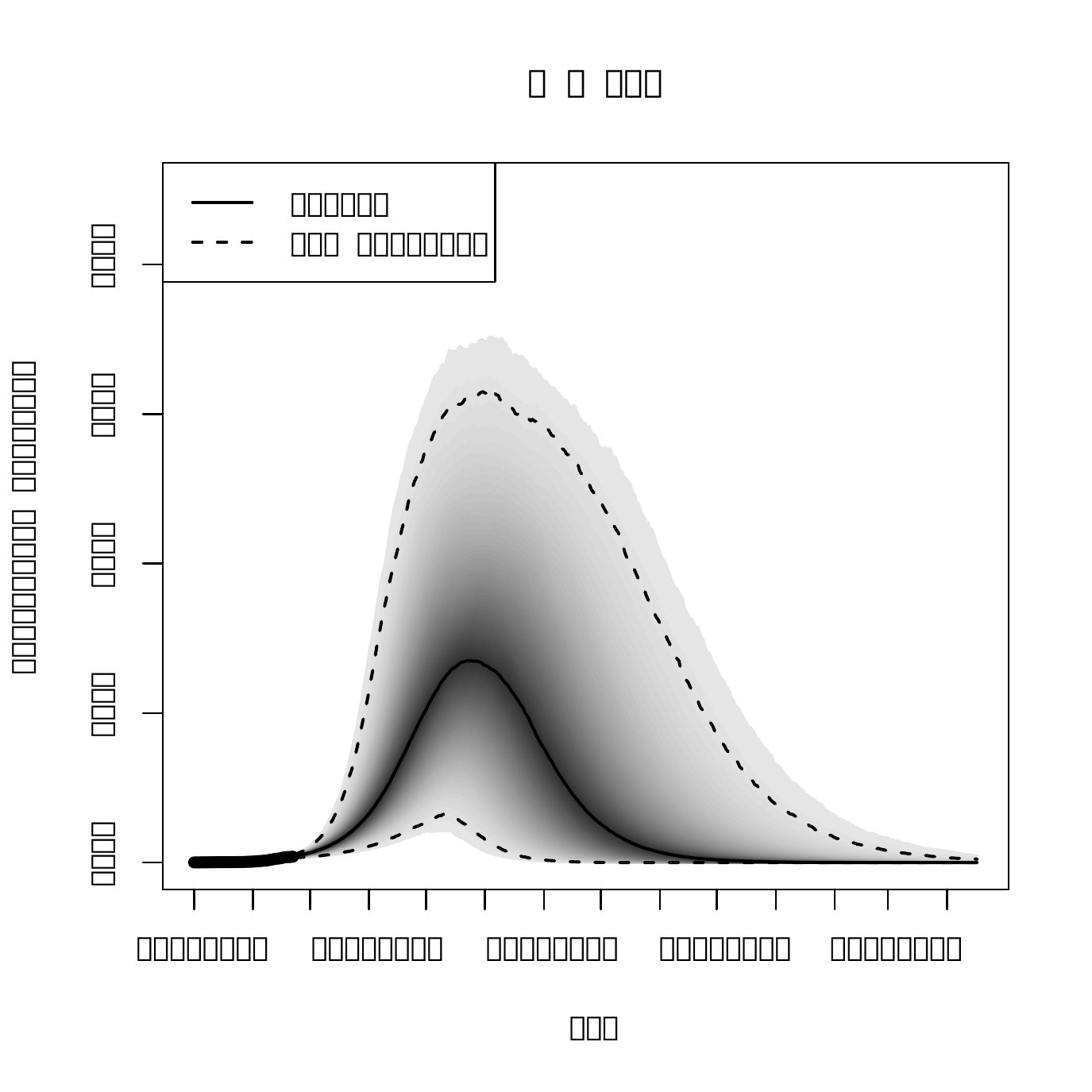}
\includegraphics[width=4.5cm]{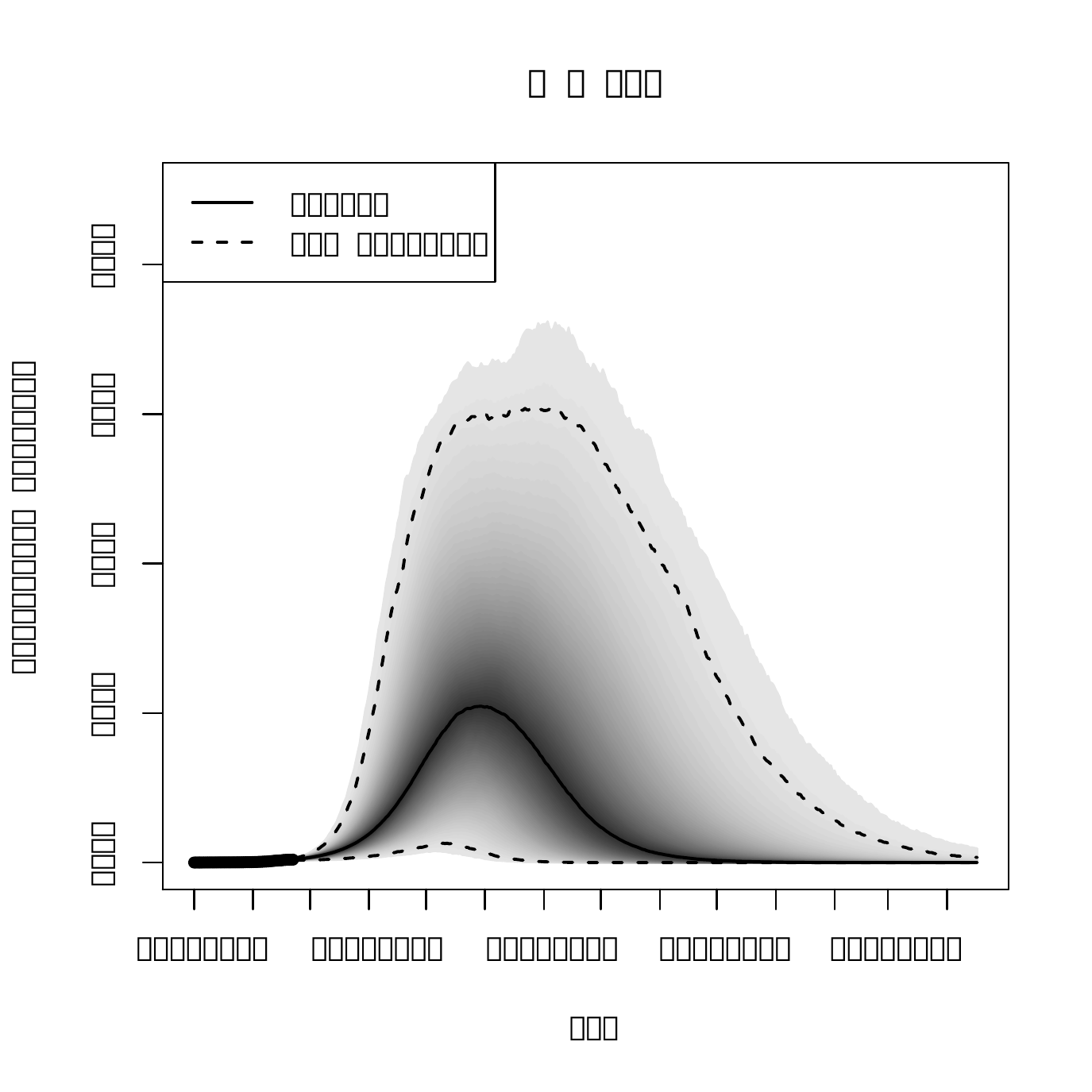}
\caption{Results of the prediction of the proportion of the infectious population with $p=0.05$ (left), $0.1$ (center) and $0.2$ (right). 
The observed data points $\{Y(t), \ t=1,\ldots, T\}$ are shown by the black dots. 
\label{fig:post}}
\end{figure}

The SS-SIR model is applied to the Japanese data with the three identification rates $p$.
First, we found that the estimates of the precision parameters $\kappa$ and $\lambda$ are very large. 
For example, the point estimates are $\widehat{\lambda}=1.75\times 10^5$ and $\widehat{\kappa}=2.64\times 10^5$  for $p=0.1$ indicating that the deterministic SIR model explains the transition of the real-time data well.
Table \ref{tab:para} reports the estimates and $95\%$ credible intervals of the representative parameters. 
%The table shows that the estimates of $\gamma$ are close $0.1$ which was adopted in the recent literature concerned with the epidemic processes of COVID-19 \cite{Kuniya2020, Sun2020}. 
%Also, the estimate of $\beta$ is somewhat close to the estimate of \cite{Kuniya2020} which was obtained by fitting the SEIR model to the similar real-time data in Japan. 
Under the three settings for $p$, the point estimates of $\beta$ are between $0.21$ and $0.25$ and those of $\gamma$ are between $0.14$ and $0.18$. 
%Those estimates $\beta$ are somewhat similar to those in \cite{Kuniya2020} found from the SEIR model applied to the early stage data of Japan. 
The estimates of the basic reproduction number $R_0$ are between $1.41$ and $1.48$. 
For $p=0.1$, for example, the 95\% credible interval of $R_0$ is (1.22--1.64). 
The estimates of PI and PT appear to vary depending on the identification rate. 
Figure \ref{fig:post} reports the future predictions of the proportion of the infectious proportion under the three identification rates. 
The figure allows us to  easily understand the degree of uncertainty in prediction, and worst and best scenarios for the future epidemic process through the interval prediction. 
It is seen that the predicted timing of the epidemic peak and peak intensity depend on the identification rate through the differences in the estimates of PT and PI. 
Specifically, the point predictions of the trajectory of the infectious proportion have the timing of the peak on July 12, July 23 and July 30 with the intensities and 95\% prediction intervals of 3.81\% (1.30\%--7.19\%), 2.7\% (0.48\%--6.23\%) and 2.09\% (0.15\%--6.04\%) for $p=0.05$, $0.1$ and $0.2$, respectively. 
The sensitivity of prediction results with respect to $p$ was also found in \cite{Kuniya2020}, but that under our setting of $p$ is far less dramatic. 
Moreover, all the scenarios predict that the epidemic peak comes during the summer 2020. 
This result is also consistent with \cite{Kuniya2020}.

\subsection{Effect of intervention}
On April 7, 2020, the Japanese government declared a state of emergency aiming at reducing human contacts by $80\%$, which is considered to be sufficient to terminate the epidemic.  
However, the government reports that the actual reduction is still limited to around $60\%$ or $70\%$ (\url{https://corona.go.jp}), mainly because the state does not have a legal force to regulate individual activities.
Also the Japanese government plans to lift the state on May 6, but the public concern lies on whether such a short period of the state of emergency is sufficient or not. 

Through simulation, we here assess the efficacy of further intervention and public awareness on mitigating the infection risk under various scenarios. 
Specifically, we consider various settings for the degrees of reduction in human contacts that are achieved by the government during the intervention and by the public awareness after the intervention, and the period of intervention denoted by $T^{\ast}$, under the state of emergency and predict the future epidemic transitions.  
Here, we focus on $p=0.1$. 
The results under $p=0.05$ and $0.2$ are found in Supplementary Material. 
It is recognized that the realization of the effect of reducing human contacts takes about two weeks since the incubation period of COVID-19 is at most 2 weeks as reported by World Health Organization.  
Since April 22, the last date in the real-time data, is almost two weeks after the declaration of the state of emergency, we assume that the infection rate changes from $\beta$ to $c\beta$ from April 23. 
For the degree of reduction in human contacts, the following six scenarios are considered: $c=0.6$, $0.5, 0.4, 0.3, 0.2$ and $0.1$
If $80\%$ reduction of human contacts was achieved, the reality would have corresponded to $c=0.2$ or $0.1$. 
In view of the current situation, however, $c=0.4$ or $0.3$ would be closer to the reality. 
We also suppose that the intervention will continue for $T^{\ast}$ days from April 23 with the three scenarios, $T^{\ast}=14, 28$ and $45$. 
Note that $T^{\ast}=14$ corresponds to May 6 on which the government is planning to lift the state. 
The other two dates to respectively correspond to the two-week and one-month extension of the intervention that continue until May 20 and June 6, respectively. 
We further suppose that the infection rate becomes $c^{\ast}\beta$ after the intervention period with the three scenarios: $c^{\ast}=1$, $0.9$ and $0.8$. 
The first scenario implies that the level of human mobility after the intervention returns to the original level before the intervention. 
The latter two scenario can reflect the remaining strain in the public awareness on mitigating the spread of infection through, for example, voluntary avoidance of outings and social distancing.

\begin{figure}[htb!]
\centering
\includegraphics[width=4.5cm]{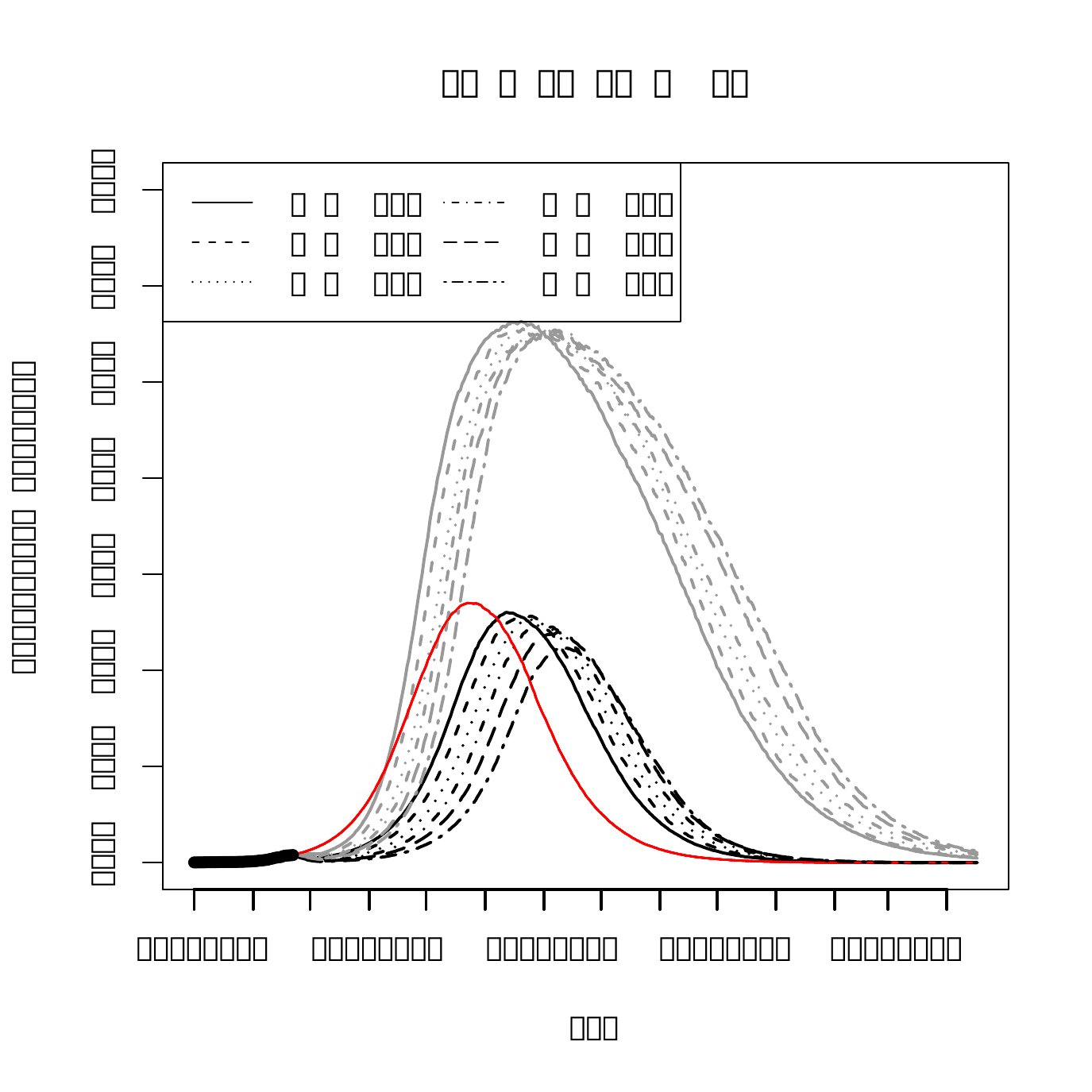}
\includegraphics[width=4.5cm]{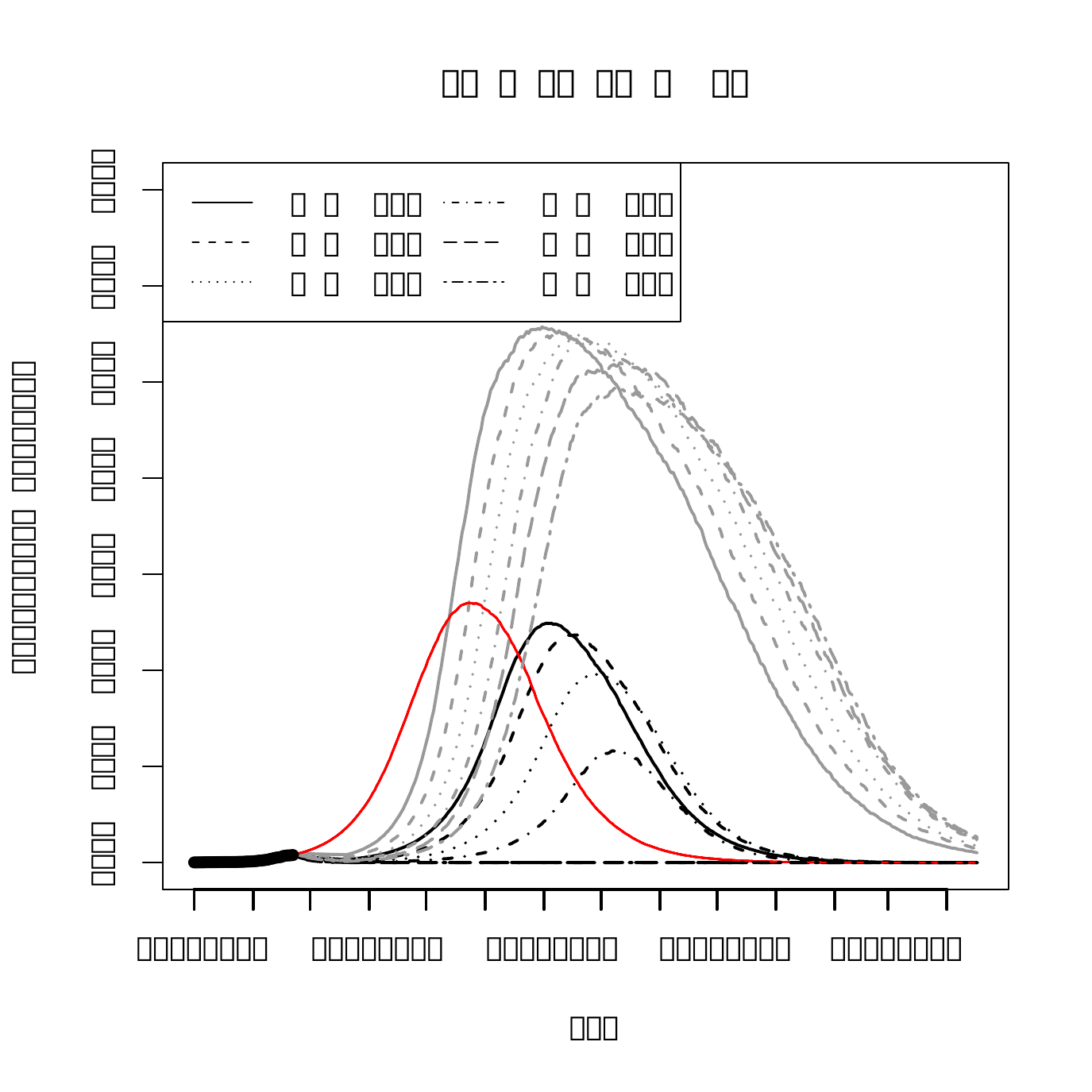}
\includegraphics[width=4.5cm]{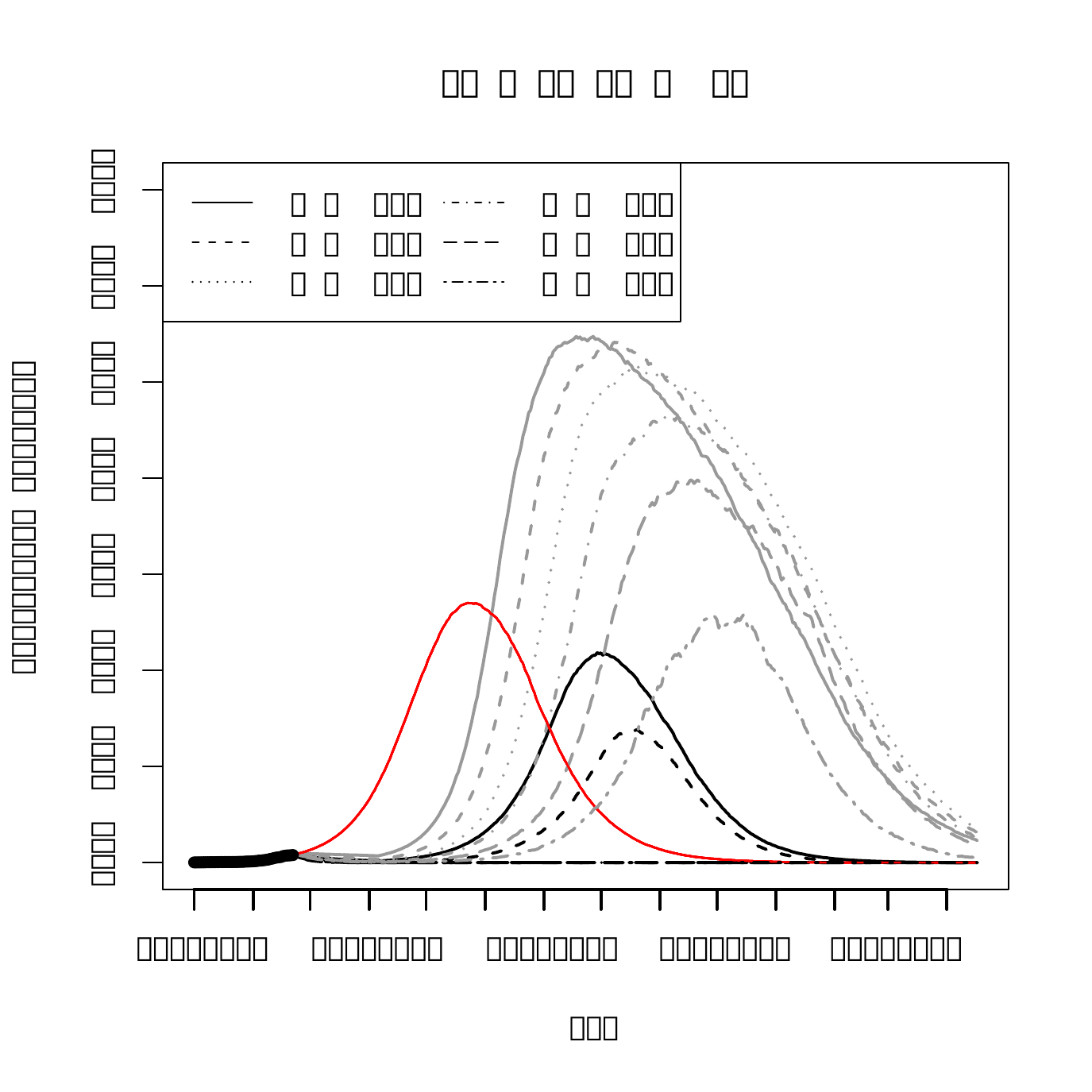}\\
\includegraphics[width=4.5cm]{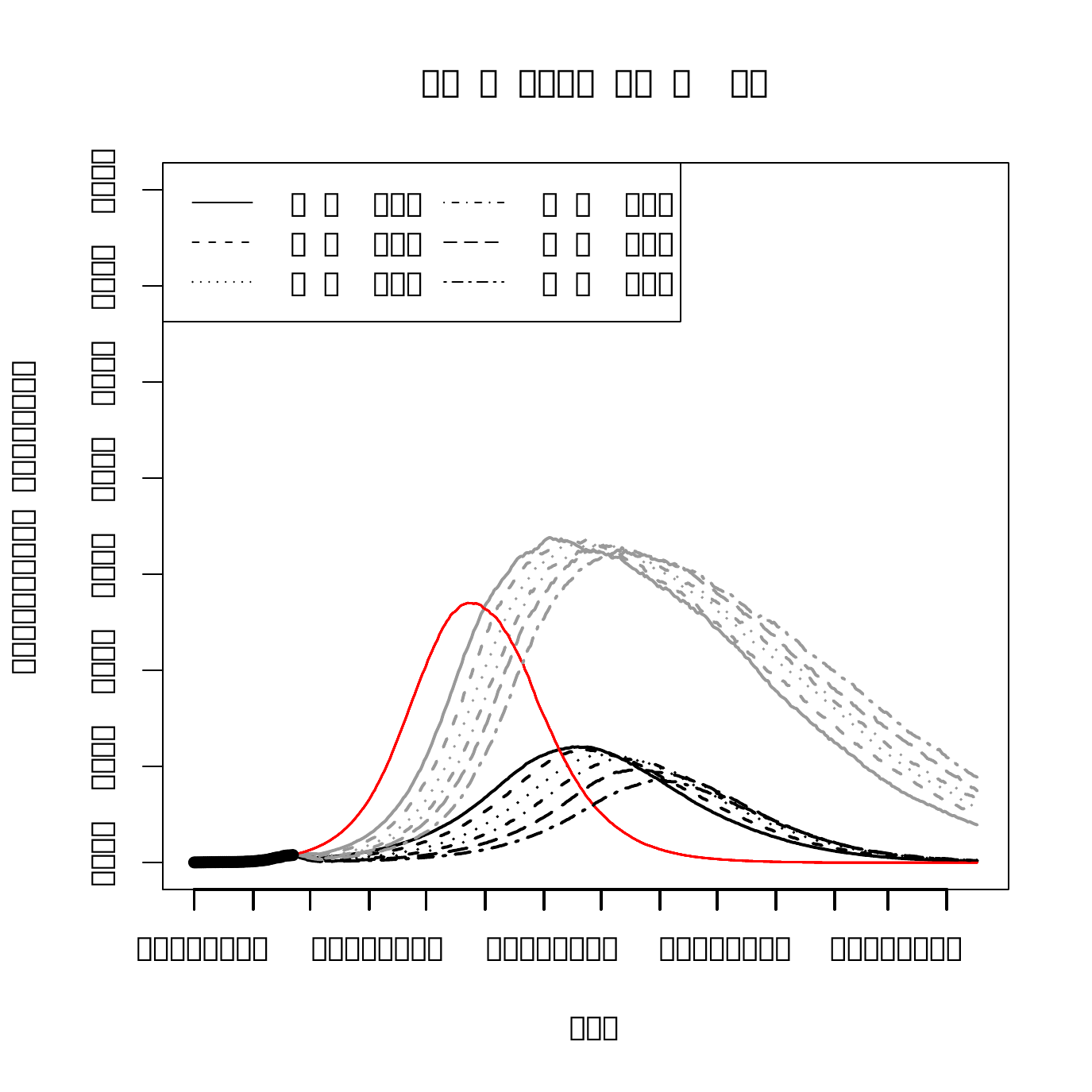}
\includegraphics[width=4.5cm]{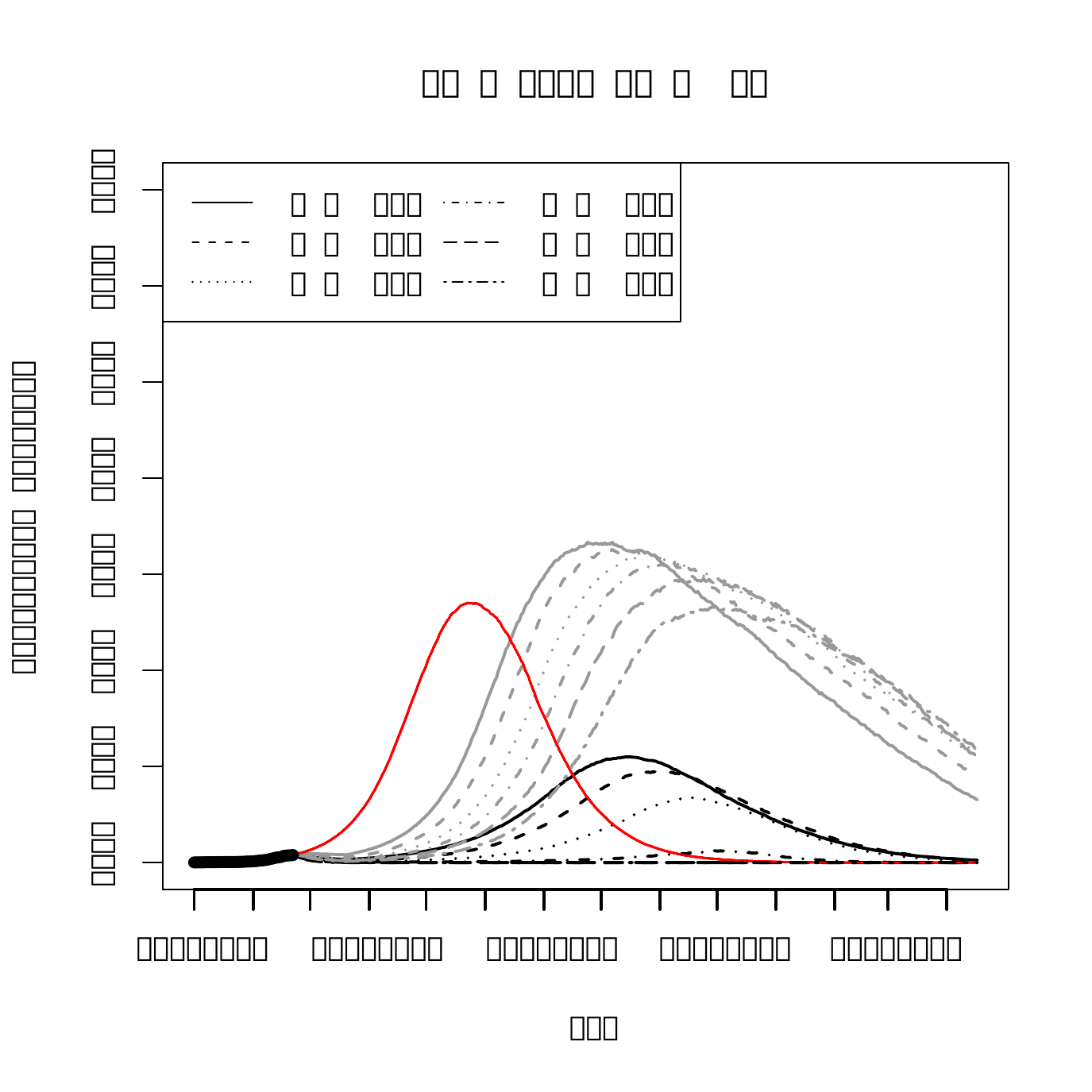}
\includegraphics[width=4.5cm]{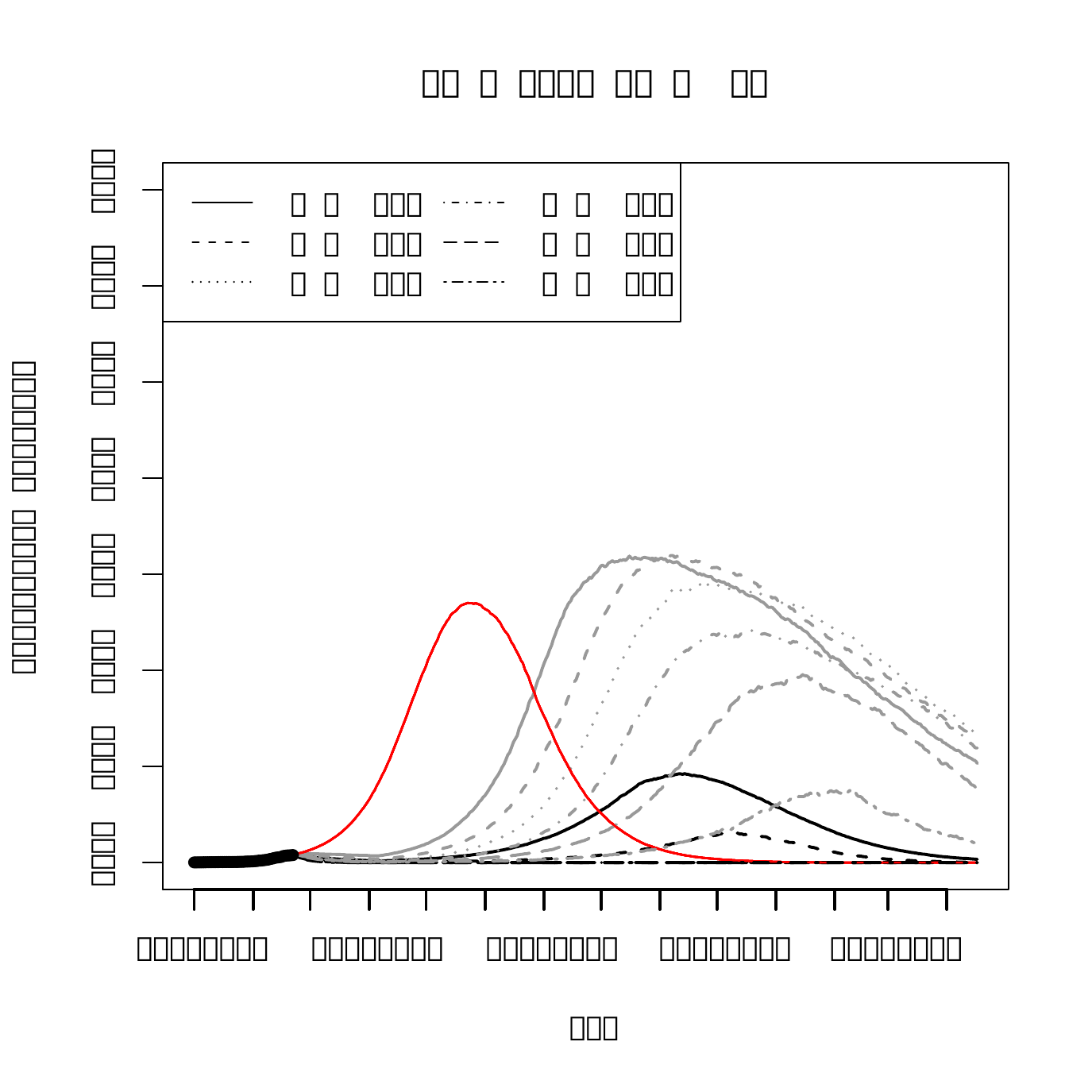}\\
\includegraphics[width=4.5cm]{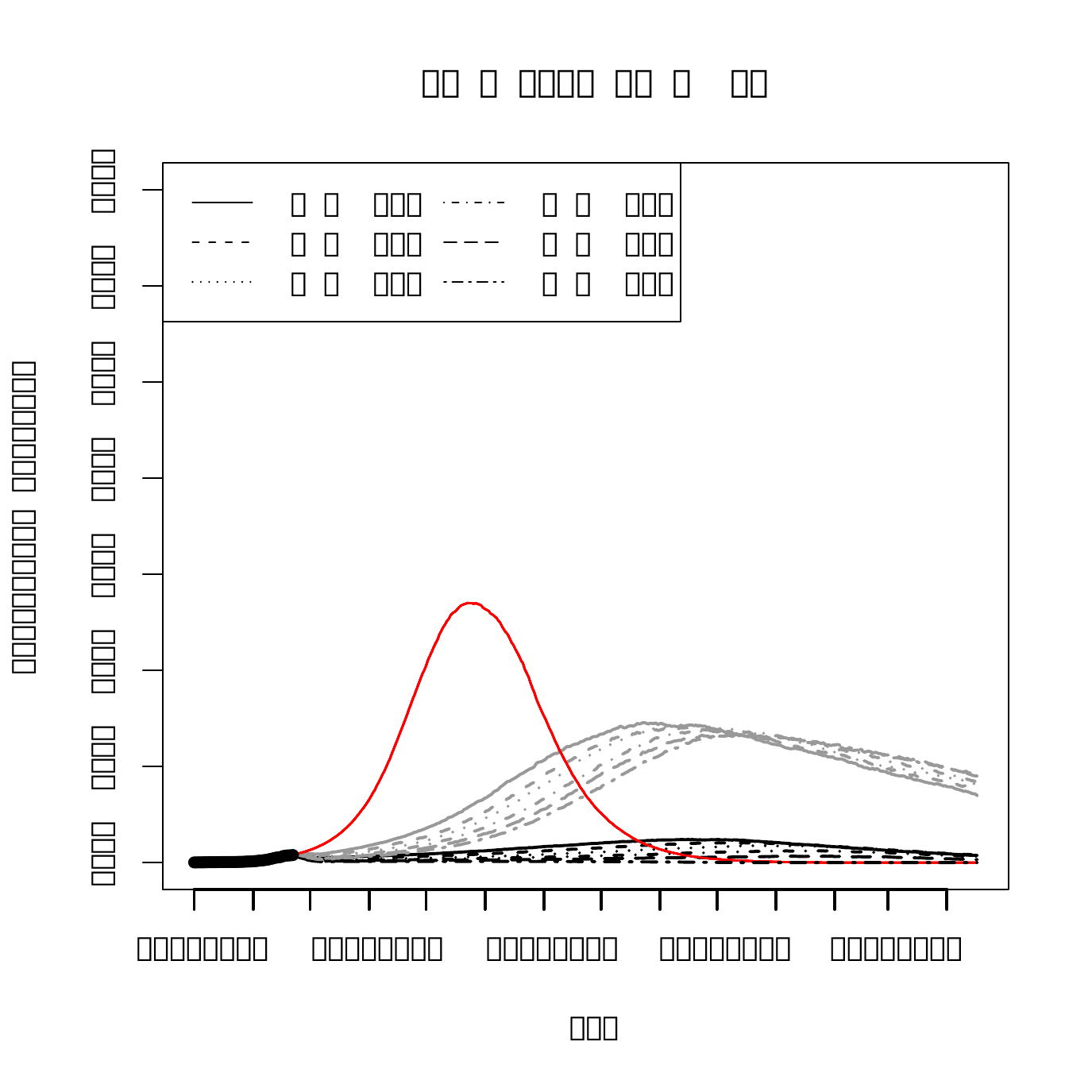}
\includegraphics[width=4.5cm]{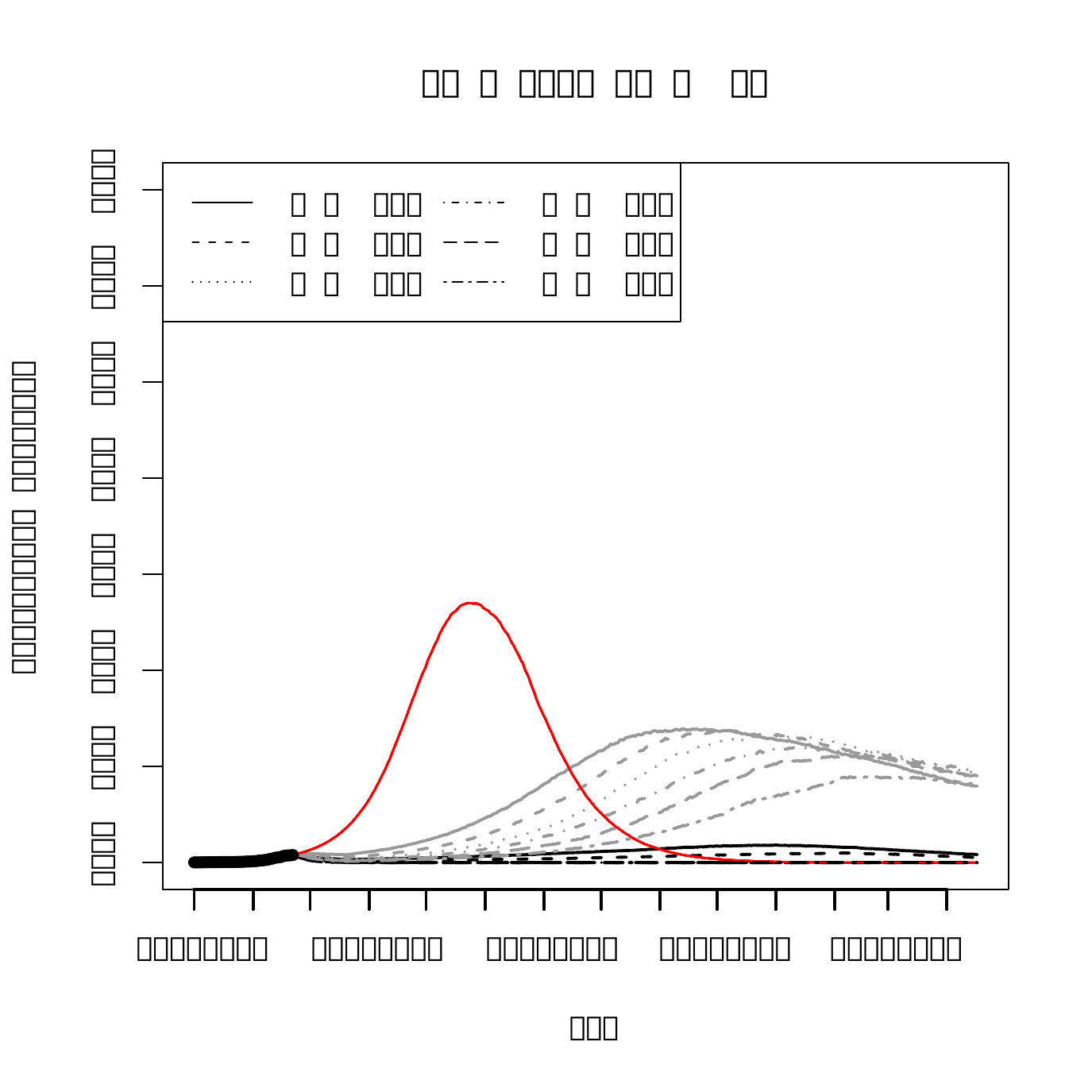}
\includegraphics[width=4.5cm]{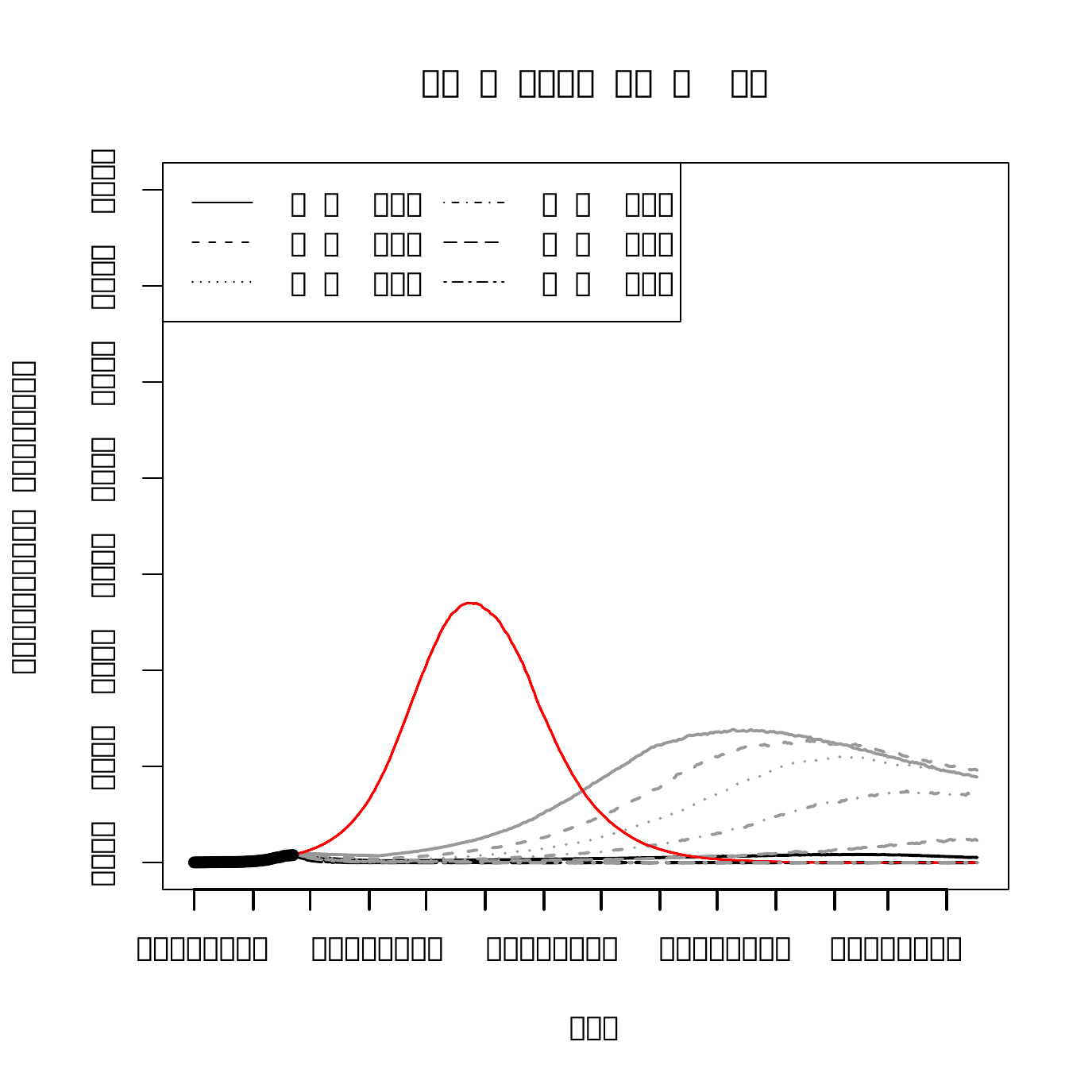}\\
\caption{Future prediction under the nine combinations of $T^{\ast}$ (the period of the intervention) and $c^{\ast}$ (the multiplier for $\beta$ after the intervention) for $p=0.1$. 
The red, black and grey curves respectively represent the future point prediction without intervention shown in Figure \ref{fig:post}, point prediction under each scenario and one-sided upper $95\%$ prediction intervals.  
\label{fig:sim}}
\end{figure}

Figure \ref{fig:sim} presents the nine panels on the future prediction under the combinations of the three scenarios of each $T^{\ast}$ and $c^{\ast}$.
Comparing the different scenarios of $T^{\ast}$, the figure reveals that setting $c$ to smaller values is effective only when it is combined with larger $T^{\ast}$. 
For example, the left upper panel of Figure \ref{fig:sim} exhibits little differences among the six choices of $c$ when $c^{\ast}=1$ and $T^{\ast}=14$.
Contrary, the small values of $c$ such as $c=0.2$ with $T^{\ast}=28$ and $45$ can lead to a convergence of the epidemic. 
Under $c=0.2$ and $c^{\ast}=1$, the epidemic can be terminated in terms of point prediction when $T^{\ast}=45$, while the epidemic peak belatedly comes on  September 8, 2020 with 2.4\% of the peak intensity when $T^{\ast}=14$. 
The result suggests that the termination of the intervention due to the currently planned lifting of the state of emergency on May 6 is too early and would only result in a slight delay in the epidemic peak and a slight reduction in the peak intensity. 

The degree of reduction in $\beta$ after the intervention, $c^{\ast}$, also has a dramatic effect on the consequence of the epidemic. 
The upper panels of Figure \ref{fig:sim} show that the efficacy of the temporary reduction in $\beta$ under the intervention can be quite limited if $\beta$ returns to the original level after the intervention. 
In contrast, if at least $20\%$ reduction in $\beta$ can be achieved for a sufficiently long period of time after the intervention, the epidemic can be effectively suppressed. 
In the case of $c^{\ast}=0.9$, for example, the peak intensity is more than halved to 1.21\% with the peak on September 16  even under the mild degree of intervention for a short period of time ($c=0.6$ and $T^{\ast}=14$). 
When a longer intervention $T^{\ast}=45$ is carried out, the peak is further delayed to November 14 with $0.93$\% of the peak intensity. 
Furthermore, in the case of $c^{\ast}=0.8$, the figure shows the epidemic is almost completely suppressed in terms of point prediction regardless of the degree of intervention and length of intervention period. 
To summarize, our results show that not only the degree of reduction in $\beta$ during the intervention but also and more importantly the length of intervention and the long term level of $\beta$ after the intervention is critical to control the spread of the epidemic.

\begin{table}[htb!]
\caption{Estimates and $95\%$ credible intervals of parameters of the SS-SIR model under  the three identification rates $p$. }
\label{tab:para}
\centering
{\footnotesize
\begin{tabular}{ccccccccccc}
\toprule
&  & \multicolumn{3}{c}{\textbf{Estimate (95\% interval)}}\\
\textbf{Parameter}	& \textbf{Description}	& \textbf{$p=0.05$} & \textbf{$p=0.1$} & \textbf{$p=0.2$} \\
\midrule
$\beta$	 & 	Infection rate	                     & 0.21  (0.13--0.34) & 0.23 (0.13--0.43)   & 0.25 (0.13--0.55)\\
$\gamma$ & 	Removal rate                         & 0.14 (0.08--0.25)  & 0.16 (0.09--0.34)   & 0.18 (0.09--0.45)\\
$R_0(=\beta/\gamma)$ & Basic reproduction number & 1.48 (1.30--1.69)  & 1.43 (1.22--1.64)   & 1.41 (1.19--1.65)\\
PT & Peak timing                                 & 145  (102--225)    & 161  (99--252)      & 157  (91--265)  \\
PI($\%$) & Peak intensity                        & 4.48 (1.80--7.80)  & 3.67 (0.99--7.02)   & 3.34 (0.68--7.19)\\
\bottomrule
\end{tabular}
}
\end{table}

%%%%%%%%%%%%%%%%%%%%%%%%%%%%%%%%%%%%%%%%%%
\section{Discussion}
In this research,  we have employed the probabilistic version of the famous SIR model, called SS-SIR model, to model the real-time data on the infectious population of COVID-19 in Japan.
The advantage of the SS-SIR model is that we can obtain not only future point prediction but also uncertainty quantification through, for example, the future prediction intervals.
The basic reproducing number $R_0$ is estimated to be approximately between 1.4 and 1.5 in this study.
This is smaller than the estimate of $2.6$ in \cite{Kuniya2020} obtained from the SEIR model applied to the early stage data in Japan. 
Note, however, that \cite{Kuniya2020} did not estimate the removal rate and onset rate but fixed their values to those found in the existing studies. 
We also estimated $R_0$ using the subset of the data up to April 6 for $p=0.1$ and the estimate is $1.44$ with the $95\%$ credible interval (1.22--1.66). 
Therefore, our estimates for $R_0$ remain unchanged even when the observations after the state of emergency are excluded. 
Moreover, our estimate in the case of Japan is also smaller than those reported from the case studies in China \citep[e.g.][]{Imai2020, Liu2020, Liu2020b, Tang2020, Wu2020}. 
Our result may have reflected the fact that the number of cases in Japan does not increase as rapidly as other countries \citep{Nippon-com}.

Through the future prediction under various scenarios on the possible reduction in the infection rate $\beta$ and the length of the intervention period, we have obtained the following epidemiological insights:
\begin{itemize}
\item
Even if a large reduction in $\beta$ could be achieved during the intervention period (e.g. the state of emergency), the convergence of the epidemic can depends on the long term value of $\beta$ after the intervention. 

\item
As long as the value of $\beta$ can be maintained to be slightly smaller even after the intervention period than that before the intervention, there is a great possibility that the epidemic terminates with a significantly smaller epidemic size than the case without intervention.

\item
The effort to reduce $\beta$ should be combined with a sufficiently long intervention period. 

\end{itemize}

These findings suggest that the lifting of the state of emergency on May 6 planned by the Japanese government will unfortunately have a too limited effect for the economic and social disturbances caused by the epidemic and state. 
A long term effort to tackle this situation is indispensable.

%%%%%%%%%%%%%%%%%%%%%%%%%%%%%%%%%%%%%%%%%%
\section*{Acknowledgement}
This research was supported by Japan Society for the Promotion of Science (KAKENHI) Grant Numbers 18K12754 and 18K12757.

\vspace{1cm}
%   Reference
%\bibliographystyle{chicago}
\bibliographystyle{unsrt}
\bibliography{refs}

\begin{thebibliography}{}

\bibitem[\protect\citeauthoryear{Bloomberg}{Bloomberg}{}]{Hokkaido}
Bloomberg.
\newblock Japan's hokkaido may have 940 infected, researcher says available
  online:
  \url{https://www.bloomberg.com/news/articles/2020-03-03/japan-s-hokkaido-could-have-up-to-940-infected-researcher-says}
  (3 march 2020).

\bibitem[\protect\citeauthoryear{Dukic, Lopes, and Polson}{Dukic
  et~al.}{2012}]{Dukic2012}
Dukic, V., H.~F. Lopes, and N.~G. Polson (2012).
\newblock Tracking epidemics with google flu trends data and a state-space seir
  model.
\newblock {\em Journal of the American Statistical Association\/}~{\em 107},
  1410--1426.

\bibitem[\protect\citeauthoryear{Gamerman and Lopes}{Gamerman and
  Lopes}{2006}]{gamerman}
Gamerman, D. and H.~F. Lopes (2006).
\newblock {\em Matkov Chain Monte Carlo: Stochastic Simulation for Bayesian
  Inference, Second Edition}.
\newblock Chapman and Hall/CRC.

\bibitem[\protect\citeauthoryear{Imai, Cori, Dorigatti, Baguelin, Connelly,
  Riley, and Ferguson}{Imai et~al.}{}]{Imai2020}
Imai, N., A.~Cori, I.~Dorigatti, M.~Baguelin, C.~Connelly, S.~Riley, and
  N.~Ferguson.
\newblock Report 3: Transmissibility of 2019-ncov; imperial college london:
  London, uk, 2020.

\bibitem[\protect\citeauthoryear{Karako, Song, Chen, and Tang}{Karako
  et~al.}{2020}]{Karako2020}
Karako, K., P.~Song, Y.~Chen, and W.~Tang (2020).
\newblock Analysis of covid-19 infection spread in japan based on stochastic
  transition model.
\newblock {\em BioScience Trends\/}, to appear.

\bibitem[\protect\citeauthoryear{Kermack and McKendrick}{Kermack and
  McKendrick}{1927}]{KK1927}
Kermack, W. and A.~McKendrick (1927).
\newblock Contribution to the mathematical theory of epidemics.
\newblock {\em Proceedings of the Royal Society of London, Series A\/}~{\em
  115}, 700--721.

\bibitem[\protect\citeauthoryear{Kuniya}{Kuniya}{2020}]{Kuniya2020}
Kuniya, T. (2020).
\newblock Prediction of the epidemic peak of coronavirus disease in japan,
  2020.
\newblock {\em Journal of Clinical Medicine\/}~{\em 9}, 789.

\bibitem[\protect\citeauthoryear{Liu, Hu, Hang, Lin, Zhong, Xiao, He, Song,
  Huang, Rong, and et~al}{Liu et~al.}{2020}]{Liu2020b}
Liu, T., J.~Hu, M.~Hang, L.~Lin, H.~Zhong, J.~Xiao, G.~He, T.~Song, Q.~Huang,
  Z.~Rong, and et~al (2020).
\newblock Transmission dynamics of 2019 novel coronavirus (2019-ncov).
\newblock {\em bioRxive\/}.

\bibitem[\protect\citeauthoryear{Liu, Gayle, Wilder-Smith, and Rocklov}{Liu
  et~al.}{2020}]{Liu2020}
Liu, Y., A.~A. Gayle, A.~Wilder-Smith, and J.~Rocklov (2020).
\newblock The reproductive number of covid-19 is higher compared to sars
  coronavirus.
\newblock {\em Journal of Travel Medicine\/}~{\em 27}, taaa021.

\bibitem[\protect\citeauthoryear{Mizumoto, Kagaya, Zarebski, and
  Chowell}{Mizumoto et~al.}{2020}]{Mizumoto2020}
Mizumoto, K., K.~Kagaya, A.~Zarebski, and G.~Chowell (2020).
\newblock Estimating the asymptomatic proportion of coronavirus disease 2019
  (covid-19) cases on board the diamond princess cruise ship, yokohama, japan,
  2020.
\newblock {\em Eurosurveillance\/}~{\em 25}.

\bibitem[\protect\citeauthoryear{Nippon.com}{Nippon.com}{}]{Nippon-com}
Nippon.com.
\newblock Coronavirus cases by country
  \url{https://www.nippon.com/en/japan-data/h00673/coronavirus-cases-by-country.html}
  (24 april 2020).

\bibitem[\protect\citeauthoryear{Osthus, Hickmann, Caragea, Higdon, and
  Del~Valle}{Osthus et~al.}{2017}]{Osthus2017}
Osthus, D., K.~Hickmann, P.~C. Caragea, D.~Higdon, and S.~Y. Del~Valle (2017).
\newblock Forecasting seasonal influenza with a state-space sir model.
\newblock {\em The Annals of Applied Statistics\/}~{\em 11}, 202--204.

\bibitem[\protect\citeauthoryear{Tang, Wang, Li, Bragazzi, Tang, Xiao, and
  Wu}{Tang et~al.}{2020}]{Tang2020}
Tang, B., X.~Wang, Q.~Li, N.~Bragazzi, S.~Tang, Y.~Xiao, and J.~Wu (2020).
\newblock Estimation of the transmission risk of the 2019-ncov and its
  implication for public health interventions.
\newblock {\em Journal of Clinical Medicine\/}~{\em 9}, 462.

\bibitem[\protect\citeauthoryear{Wu, Leung, and Leung}{Wu
  et~al.}{2020}]{Wu2020}
Wu, J., K.~Leung, and G.~Leung (2020).
\newblock Nowcasting and forecasting the potential domestic and international
  spread of the 2019-ncov outbreak originating in wuhan, china: A modelling
  study.
\newblock {\em Lancet\/}~{\em 395}, 689--697.

\end{thebibliography}

\newpage
%--------------------------------------------------------------------------
%    Supplement 
%--------------------------------------------------------------------------
\begin{center}
{\LARGE 
{\bf Supplementary Material for ``Predicting Infection of COVID-19 in Japan: State Space Modeling Approach" }
}
\end{center}

%  new environment
\setcounter{equation}{0}
\renewcommand{\theequation}{S\arabic{equation}}
\setcounter{section}{0}
\renewcommand{\thesection}{S\arabic{section}}
\setcounter{lem}{0}
\renewcommand{\thelem}{S\arabic{lem}}
\setcounter{page}{1}
\setcounter{figure}{0}
\renewcommand{\thefigure}{S\arabic{figure}}

\vspace{0.5cm}
We here provide details of the prior distributions and Markov Chain Monte Carlo algorithm to generate posterior samples.

\section{Model}
The state space SIR (SS-SIR) model consists of the following two equations: 
\begin{equation}\label{model-S}
\begin{split}
&Y(t)|I(t)\sim {\rm Beta}(\lambda I(t), \lambda(1-I(t))), \\
&\theta(t)|\theta(t-1) \sim {\rm Dir}(\kappa f(\theta(t-1);\beta, \gamma)),
\end{split}
\end{equation}
for $t=1,\ldots,T$ with $T=53$ in our analysis, where $\theta(t)=(S(t), I(t), R(t))$, $\beta>0$ and $\gamma>0$ are the unknown infection rate and removal rate, respectively, and $f(\theta(t-1);\beta, \gamma)$ is the solution of the deterministic SIR model starting at $\theta(t-1)$:  
\begin{equation*}
S'(t)=-\beta S(t)I(t), \ \ \ \ \ I'(t)=\beta S(t)I(t) -\gamma I(t), \ \ \ \ \ R'(t)=\gamma I(t). 
\end{equation*}

\section{Prior distributions}\label{sec:prior}

\subsection{Precision parameters $\kappa$ and $\lambda$}  
We set $\kappa\sim {\rm Ga}(20, 0.0001)$ and $\lambda\sim {\rm Ga}(2,0.0001)$, independently.

\subsection{Initial state $\theta(0)=(S(0),I(0), R(0))$}
The joint prior is constructed via the following decomposition: 
$$
\pi(S(0),I(0), R(0))=\pi(S(0))\pi(I(0))\pi(R(0)|S(0),I(0)). 
$$
The prior distributions of  $S(0)$ and $I(0)$ are first determined and then the prior distribution of $R(0)$ is determined accordingly. 
We assume that $95\%$ of population is initially susceptible, $S(0)\sim \delta({0.95})$, and $I(0)$ follows a beta distribution. 
The parameters of the beta distribution is set such that $E[I(0)]=1.5\time 10^{-4}$ for $p=0.05$ and $E[I(0)]=8.0\times10^{-5}$ for $p=0.1$ and $0.2$, and ${\rm Var}(I_0)=1.0\times 10^{-8}$ for all the cases.

\subsection{Key parameters $\beta$ and $\gamma$ in the SIR model}
The prior distributions of the two important parameters $\beta$ and $\gamma$ through the prior distributions of $\rho=\gamma/\beta$,  peak intensity (PI) and timing of the peak intensity (PT). 
The prior distribution of $\rho$, reciprocal of {\it the basic reproduction number}, is firstly derived.
If $\rho\in [0, S(0)]$, $I(t)$ starts increasing, reaches its maximum  and decreases to zero as $t\xrightarrow \infty$ so the model is designed as epidemic.  
Under the SIR model, PI can be expressed as 
$$
{\rm PI}\equiv g(S(0),I(0),\rho)=I(0)+S(0)-\rho(\log(S(0))+1-\log(\rho)),
$$
and it is known that the unique solution $\rho=g^{-1}({\rm PI},S(0),I(0))$ exists.
Then, the prior distribution of $\rho$ is determined by specifying the prior distribution of PI. 
We assume ${\rm PI}\sim {\rm TN}_{(I(0),1)}(0.03,0.02^2)$, that is, the prior peak intensity is centered around 3\% of the population. 
Similarly, the prior distribution of PT is given by ${\rm PT}\sim {\rm TN}_{(53,413)}(180, 60^2)$, implying the prior timing of the epidemic peak is between April 22, 2020 and April 17, 2021, and is centered around August 27, 2020.

Next, the prior distribution of $\beta$ is specified following the regression approach of \cite{Osthus2017}.
We first prepare the grids on the space of  $\beta$, PI and $I(0)$ as follow: 
\begin{itemize}
\item
$\beta$: equally spaced 40 points between $0.05$ and $1.0$,

\item
PI:  points equally spaced between $0.01$ and $0.1$ by $0.01$,

\item
I(0): equally spaced 20 points between $0.1\times Y(1)$ and $0.001$. 
\end{itemize} 
The SIR curves are simulated for all the combinations of $(\beta, {\rm PI}, I(0))$ and PT is identified. 
Then $\log(\beta)$ is regressed on a subset of a fourth degree polynomial interaction model using $\log({\rm PT})$, $\log I(0)$, and $\log(\rho)$ as covariates. 
Since our grid for $\beta$ covers much smaller values than \cite{Osthus2017}, our regression model contains 28 covariates in total including the constant and they are collectively denoted by $X$. 
The regression coefficients and error variance are denoted by $\hat{\tau}$ $\hat{\sigma}^2$, respectively. 
Based on these estimates, we set $\beta\sim \delta(\exp(X\hat{\tau}+0.5\hat{\sigma}^2))$.
The list of the used covariates and estimate of the polynomial regression are presented in Table \ref{tab:reg_fit}. 
Also, $\hat{\sigma}^2=6.92\times10^{-8}$ and $R^2$ is almost equal to one. 
 
\begin{table}[htbp]
    \centering
    \begin{tabular}{lcll|lcll}\toprule
Covariate & $\tau$ & Estimate & Std. Error &Covariate & $\tau$ & Estimate & Std. Error  \\\hline
Intercept                          & $\tau_{1}$  & -2.45                &  2.28$\times10^{-1}$ &$I(0)^2\times\rho^3$               & $\tau_{15}$ & -7.22$\times10^{-1}$ &  1.16$\times10^{-1}$ \\
$\rm{PT}$                          & $\tau_{2}$  & -8.55$\times10^{-1}$ &  4.02$\times10^{-2}$ &$I(0)\times\rho^4$                 & $\tau_{16}$ & -1.16$\times10$      &  1.97                \\
$\rm{PT}^2$                        & $\tau_{3}$  & -6.20$\times10^{-2}$ &  1.37$\times10^{-2}$ &$I(0)^2\times\rho^4$               & $\tau_{17}$ & -3.77$\times10^{-1}$ &  7.23$\times10^{-2}$ \\
$I(0)$                             & $\tau_{4}$  & -1.20                &  4.10$\times10^{-2}$ &$\rm{PT}\times I(0)$               & $\tau_{18}$ & -1.78$\times10^{-2}$ &  4.42$\times10^{-3}$ \\
$I(0)^2$                           & $\tau_{5}$  & -5.14$\times10^{-2}$ &  1.57$\times10^{-3}$ &$\rm{PT}\times \rho$               & $\tau_{19}$ &  2.98$\times10^{-2}$ &  1.07$\times10^{-2}$ \\
$\rho$                             & $\tau_{6}$  & -2.58$\times10^{ 1}$ &  2.52                &$\rm{PT}^2\times I(0)$             & $\tau_{20}$ &  2.51$\times10^{-3}$ &  8.68$\times10^{-4}$ \\
$\rho^2$                           & $\tau_{7}$  & -8.23$\times10^{ 1}$ &  1.02$\times10$      &$\rm{PT}\times I(0)^2$             & $\tau_{21}$ & -4.95$\times10^{-4}$ &  1.93$\times10^{-4}$ \\
$\rho^3$                           & $\tau_{8}$  & -1.16$\times10^{ 2}$ &  1.76$\times10$      &$\rm{PT}^2\times \rho$             & $\tau_{22}$ & -2.16$\times10^{-3}$ &  1.19$\times10^{-3}$ \\
$\rho^4$                           & $\tau_{9}$  & -6.19$\times10^{ 1}$ &  1.09$\times10$      &$\rm{PT}^2\times \rho^2$           & $\tau_{23}$ &  1.13$\times10^{-3}$ &  1.03$\times10^{-3}$ \\
$I(0)\times\rho$                   & $\tau_{10}$ & -5.52                &  4.55$\times10^{-1}$ &$\rm{PT}^3$                        & $\tau_{24}$ &  1.01$\times10^{-2}$ &  2.33$\times10^{-3}$ \\
$I(0)^2\times\rho$                 & $\tau_{11}$ & -1.76$\times10^{-1}$ &  1.67$\times10^{-2}$ &$\rm{PT}^4$                        & $\tau_{25}$ & -5.61$\times10^{-4}$ &  1.51$\times10^{-4}$ \\
$I(0)\times\rho^2$                 & $\tau_{12}$ & -1.62$\times10$      &  1.84                &$\rm{PT}^3\times I(0)$             & $\tau_{26}$ & -9.43$\times10^{-5}$ &  7.19$\times10^{-5}$ \\
$I(0)^2\times\rho^2$               & $\tau_{13}$ & -5.23$\times10^{-1}$ &  6.76$\times10^{-2}$ &$\rm{PT}^2\times I(0)^2$           & $\tau_{27}$ &  4.93$\times10^{-5}$ &  2.04$\times10^{-5}$ \\
$I(0)\times\rho^3$                 & $\tau_{14}$ & -2.23$\times10$      &  3.17                &$I(0)^3$                           & $\tau_{28}$ & -6.46$\times10^{-4}$ &  1.21$\times10^{-5}$  \\
\bottomrule
    \end{tabular}
    \caption{Regression estimates for the prior distribution of $\beta$}
    \label{tab:reg_fit}
\end{table}

\subsection{Prior predictive distribution}
\begin{figure}[htbp]
\centering
\includegraphics[width=4.5cm]{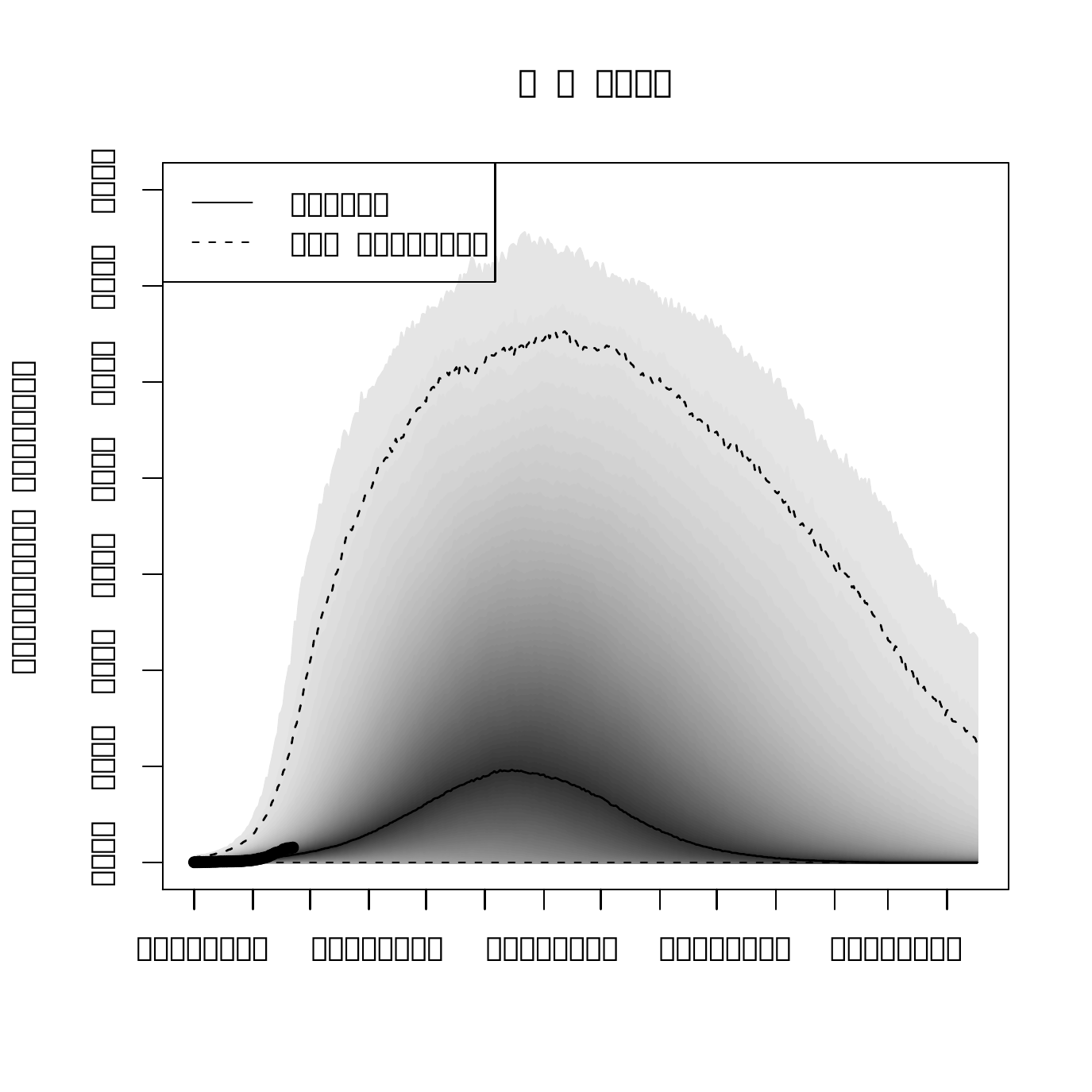}
\includegraphics[width=4.5cm]{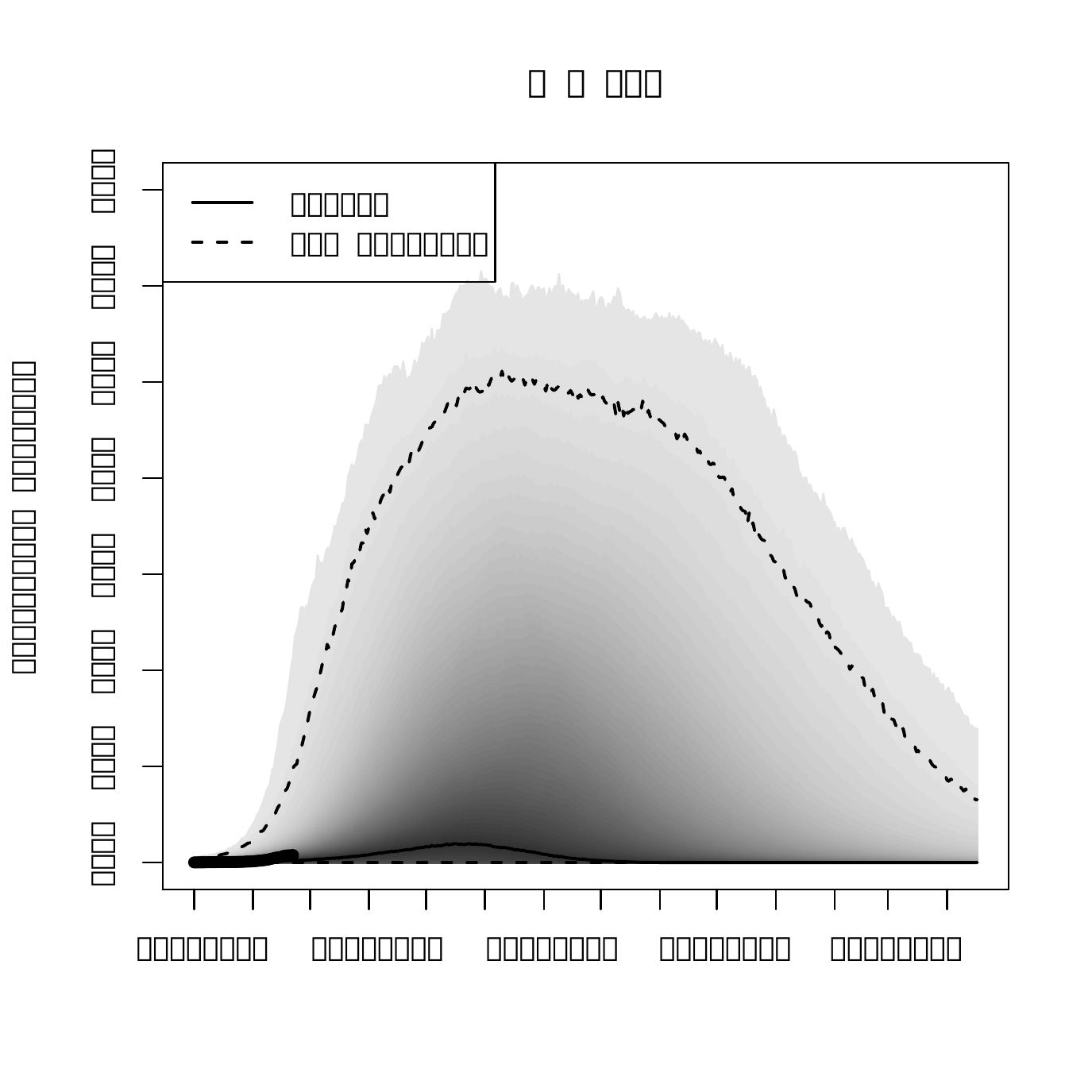}
\includegraphics[width=4.5cm]{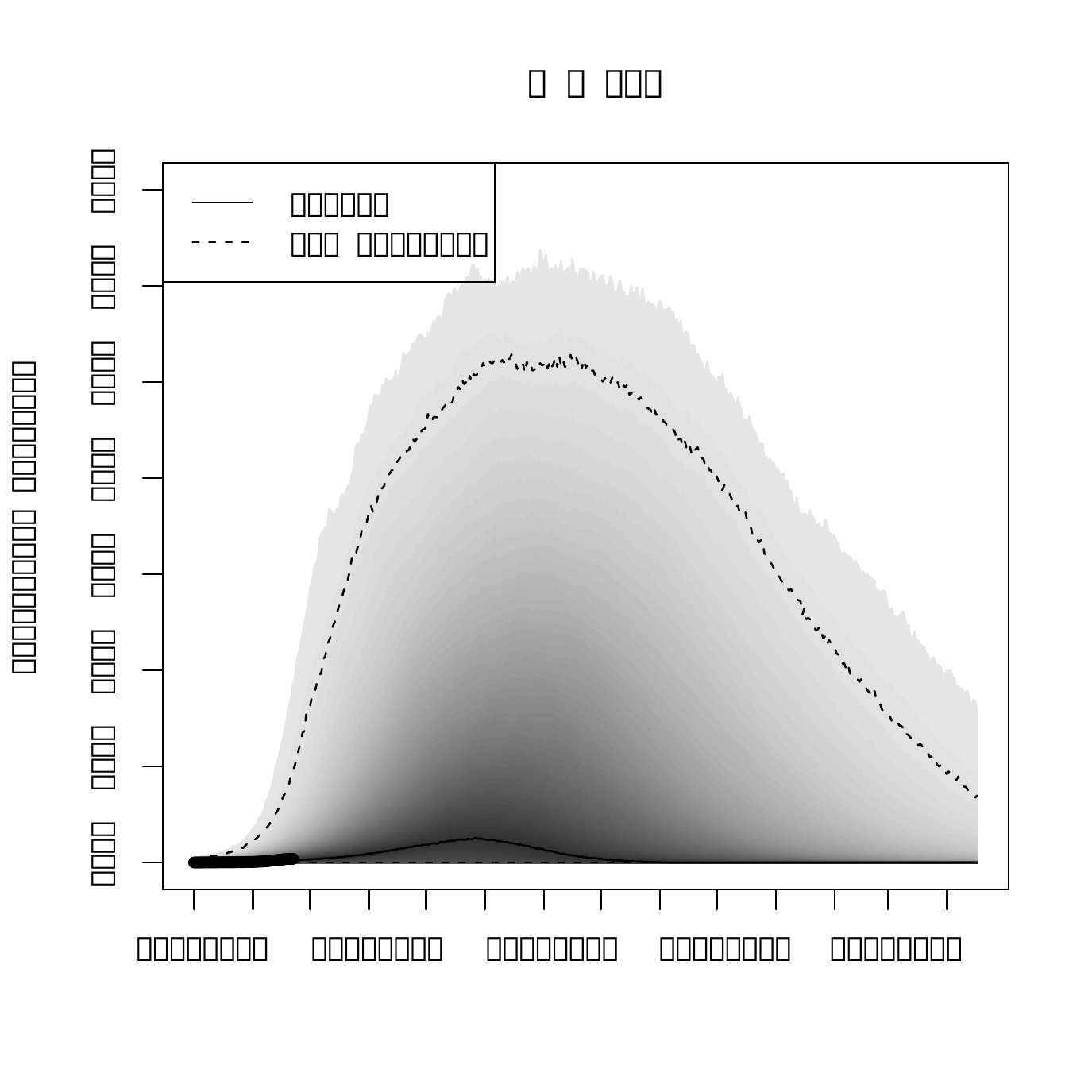}
\caption{Prior predictive distributions for $p=0.05$ (left), $0.1$  and $0.2$ (right).
The observed data points $\{Y(t), \ t=1,\ldots, T\}$ are plotted by the black dots. \label{fig:prior}}
\end{figure}

Under the prior distributions specified as above, we generated random samples from the {\it prior predictive distribution} given by 
$$
\pi(Y_{1:T})=\int\int \prod_{t=1}^T\left[ f(Y(t)|\theta(t),\phi)g(\theta(t)|\theta(t-1),\phi)\right]\pi(\phi)d \theta_{1:T}d\phi,
$$
where $Y_{1:T}=(Y(1),\ldots, Y(T))$ and $\theta_{1:T}=(\theta(1),\ldots,\theta(T))$. 
The prior predictive distribution presented in Figure \ref{fig:prior} shows that the observed data are included in the prior prediction intervals indicating that our settings of the prior distributions are reasonable.

\section{Posterior distribution and Markov Chain Monte Carlo sampling algorithm}
Let $\phi$ denote the collection of the unknown parameters, $\phi=(\beta,\rho,\kappa,\lambda,I(0))$, and $\pi(\phi)$ denote the joint prior distribution specified in Section \ref{sec:prior}. 
The posterior distribution of the latent variables $\theta(t)$ and $\phi$ is given by 
\begin{equation}\label{pos-S}
\pi(\theta_{1:T},\phi|Y_{1:T})\propto\prod_{t=1}^{T}\left[ f(Y(t)|\theta(t);\phi)g(\theta(t)|\theta(t-1);\phi)\right] \pi(\phi),
\end{equation}
where $f(\cdot)$ and $g(\cdot)$ are the conditional distribution of the first and second equations in the model (\ref{model-S}), respectively. 
Since the posterior distribution (\ref{pos-S}) has a complicated form, the posterior inference is based on the Markov Chain Monte Carlo (MCMC) sampling method.
Specifically, we adopt the Metropolis-Hastings (MH) within Gibbs sampling, in which the random numbers are alternately sampled from the full conditional distributions $\phi$ and $\theta_{1:T}$. 
Our MCMC algorithm repeats the following steps:
\begin{itemize}
\item[-]
Sample from $\theta(t)|\theta(t+1), \theta(t-1), \phi, Y(t)$ for $t=1,\ldots,T$.

\item[-]
Sample from $I(0), \rm{PI}, \rm{PT}, \kappa|\theta_{1:T}, \lambda, Y_{1:T}$.

\item[-]
Sample from $\lambda|\theta_{1:T}, Y_{1:T}$.

\end{itemize} 

In each step, the sampling is carried out by using the Gaussian random walk MH algorithm where the step sizes are adjusted such that the acceptance rates are between 0.2 and 0.4. 
Note that given the sampled values of $I(0)$, PI and PT, the values of $\rho$, $\beta$ and hence $\gamma=\beta\rho$ and $R_0=1/\rho$ are immediately determined through their prior distributions. 
We run the algorithm for  50000 iterations after 10000 iterations of initial burn-in period. 
Then every 10th draws (5000 draws) is retained to be used in our analysis and the point estimates and 95\% credible intervals are the sample medians and $0.025$th and $0.975$th sample quantiles of the MCMC output.

To carry out future predictions for the periods $t=T+1,\dots,T^{'}$, we generate random numbers from the {\it posterior predictive distribution} given by
$$
\pi(Y_{T+1:T^{'}}|Y_{1:T})=\int\int \prod_{t={T+1}}^{T^{'}}\left[ f(Y(t)|\theta(t),\phi)g(\theta(t)|\theta(t-1),\phi)\right]\pi(\phi,\theta_{1:T}|Y_{1:T})d \theta_{1:T^{'}}d\phi. 
$$
Given the MCMC outputs and values of $c$, $T^{\ast}$ and $c^{\ast}$ as specified in our prediction scenarios, the future prediction of the epidemic is carried out by repeating the following steps for $t=T+1,\dots,T^{'}$:
    \begin{itemize}
        \item 
        If $t\leq T^{\ast}$, sample from ${\rm Dir}(\kappa f(\theta(t-1); c\beta, \gamma)$, otherwise sample $\theta(t)$ from ${\rm Dir}(\kappa f(\theta(t-1); c^{\ast}\beta, \gamma)$.
        \item
        sample $Y(t)$ from ${\rm Beta}(\lambda I(t), \lambda(1-I(t)))$
    \end{itemize}

Our point prediction of the trajectory of the infectious proportion is obtained from the sample medians of the simulated $Y(t)$ computed at each $t$. 
Similarly, the prediction interval is obtained from the $0.025$th and $0.975$th sample quantiles.

\section{Additional results}
\subsection{MCMC and posterior distribution of $I(t)$}
\begin{figure}[htbp]
\centering
\includegraphics[width=10cm]{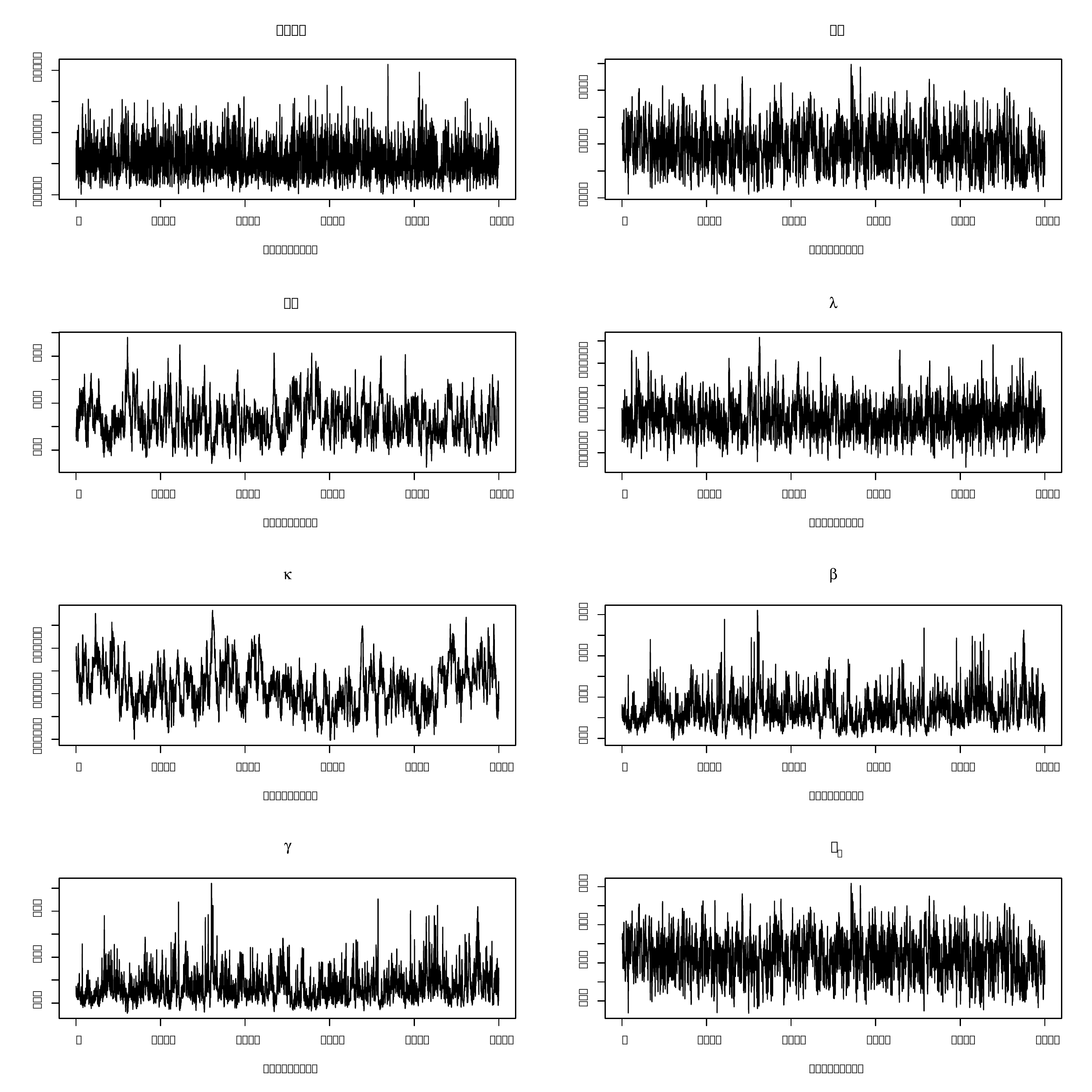}
\caption{Trace plots of the MCMC output ($p=0.1$).
 \label{fig:trace}}
\end{figure}

\begin{figure}[htbp]
\centering
\includegraphics[width=8cm]{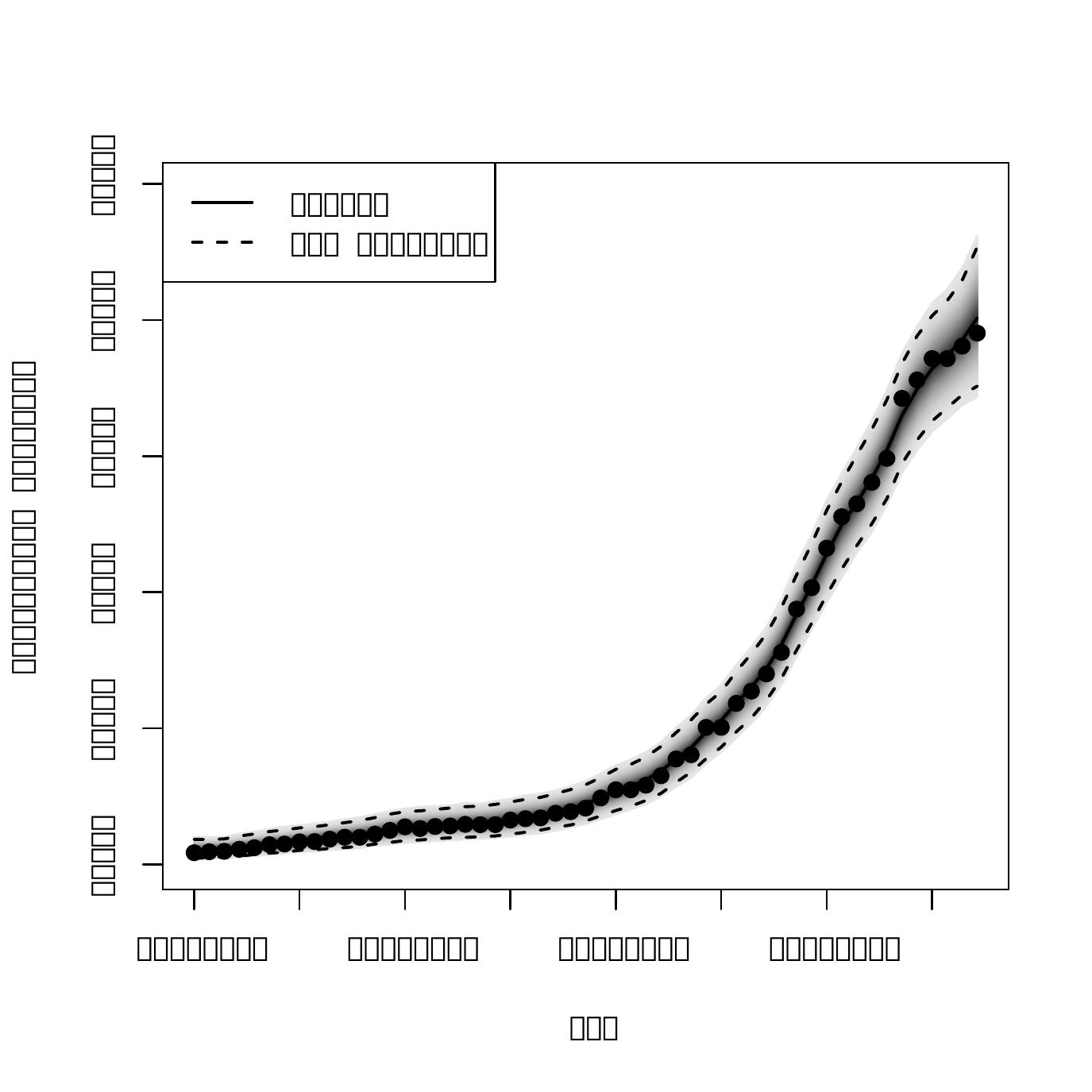}
\caption{Posterior distribution and 95\% credible interval of $I(t)$ ($p=0.1$). The observed data points are shown by the black dots.
 \label{fig:ci}}
\end{figure}

Figure \ref{fig:trace} presents the trace plots of the MCMC output for $p=0.1$. 
The figure shows that our MCMC algorithm converges to the target distributions and mixes reasonably well. 
Figure \ref{fig:ci} presents the posterior distribution of $I(t)$ for $t=1,\dots,53$. 
It is shown that the observed data points are well within  the 95\% credible interval.

\subsection{Additional prediction results for $p=0.1$}
\begin{figure}[htbp]
\centering
\includegraphics[width=4.6cm]{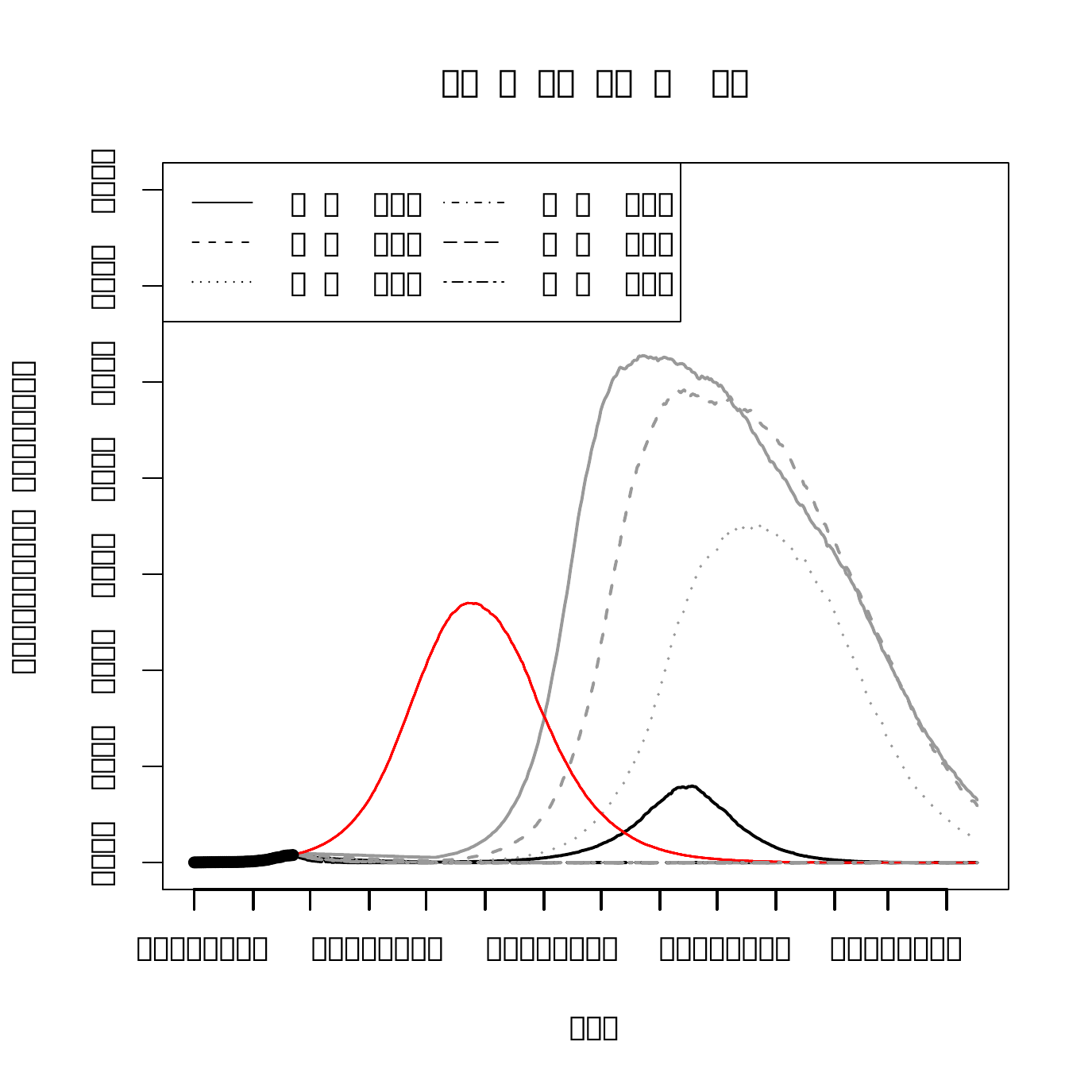}
\includegraphics[width=4.6cm]{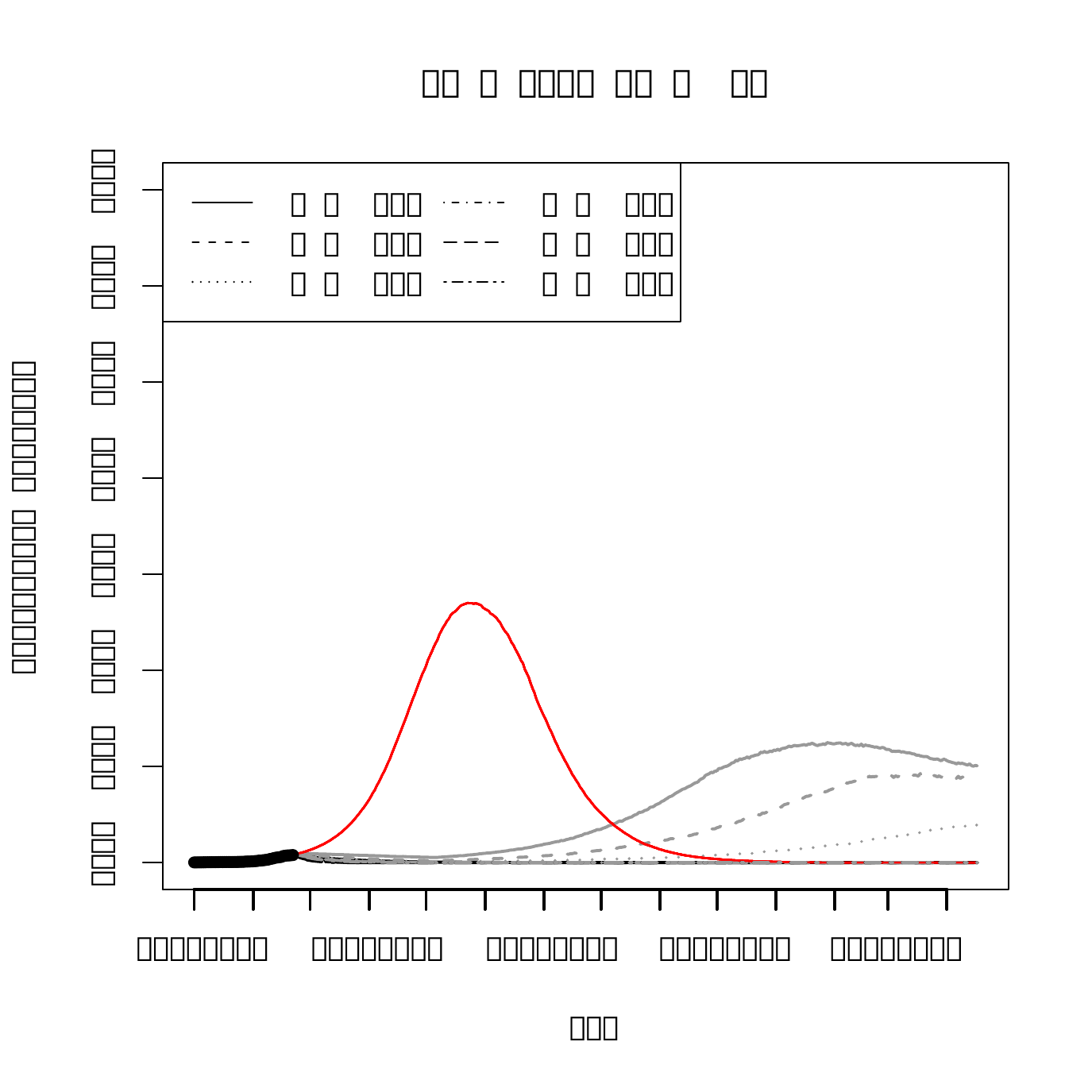}
\includegraphics[width=4.6cm]{fig_pred2_s3_4.pdf}
\caption{Future prediction with $T^{\ast}=75$ (75 days of intervention after April 22) for $c^{\ast}=1$ (left), $0.9$ (center) and $0.8$ (right). The red, black and grey curves represent the future point prediction without intervention shown in Figure 2 in the main text, point prediction under each scenario and one-sided upper 95\% prediction intervals, respectively.
 \label{fig:add-sim1}}
\end{figure}

In Section 3 of the main text, we assessed the effect of an intervention based on the future prediction under several scenarios. 
Here the additional results under a longer intervention period with $T^{\ast}=75$ are provided in Figure~\ref{fig:add-sim1}. 
The figure shows that even if $\beta$ returns to the original level the intervention ($c^{\ast}=1$), as long as the intervention period is sufficiently long, even a somewhat mild but realistic degree of intervention, such as $c=0.4$, can lead to the termination of the epidemic. 
     
\subsection{Prediction results for $p=0.05$ and $0.2$}
\begin{figure}[htb!]
\centering
\includegraphics[width=4.6cm]{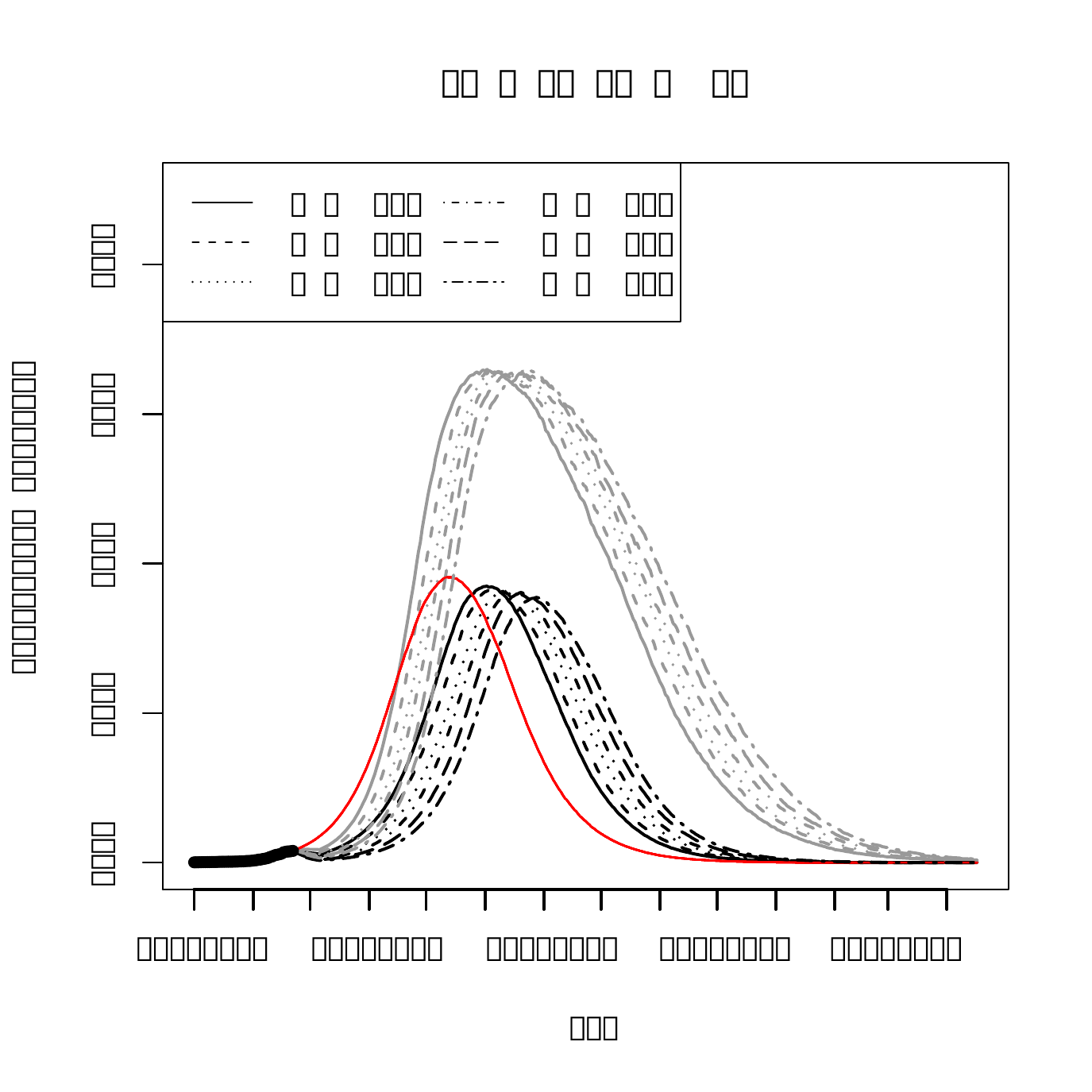}
\includegraphics[width=4.6cm]{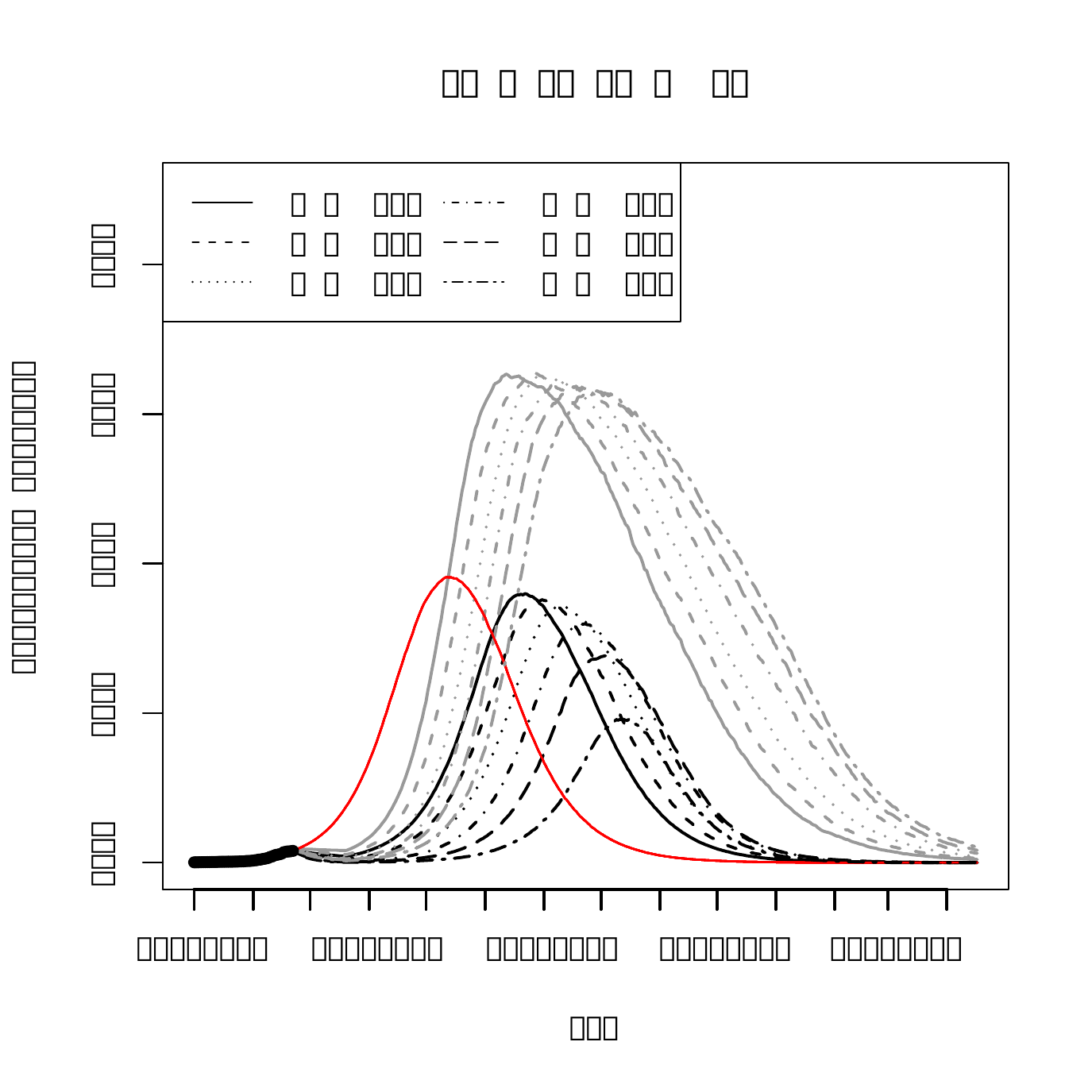}
\includegraphics[width=4.6cm]{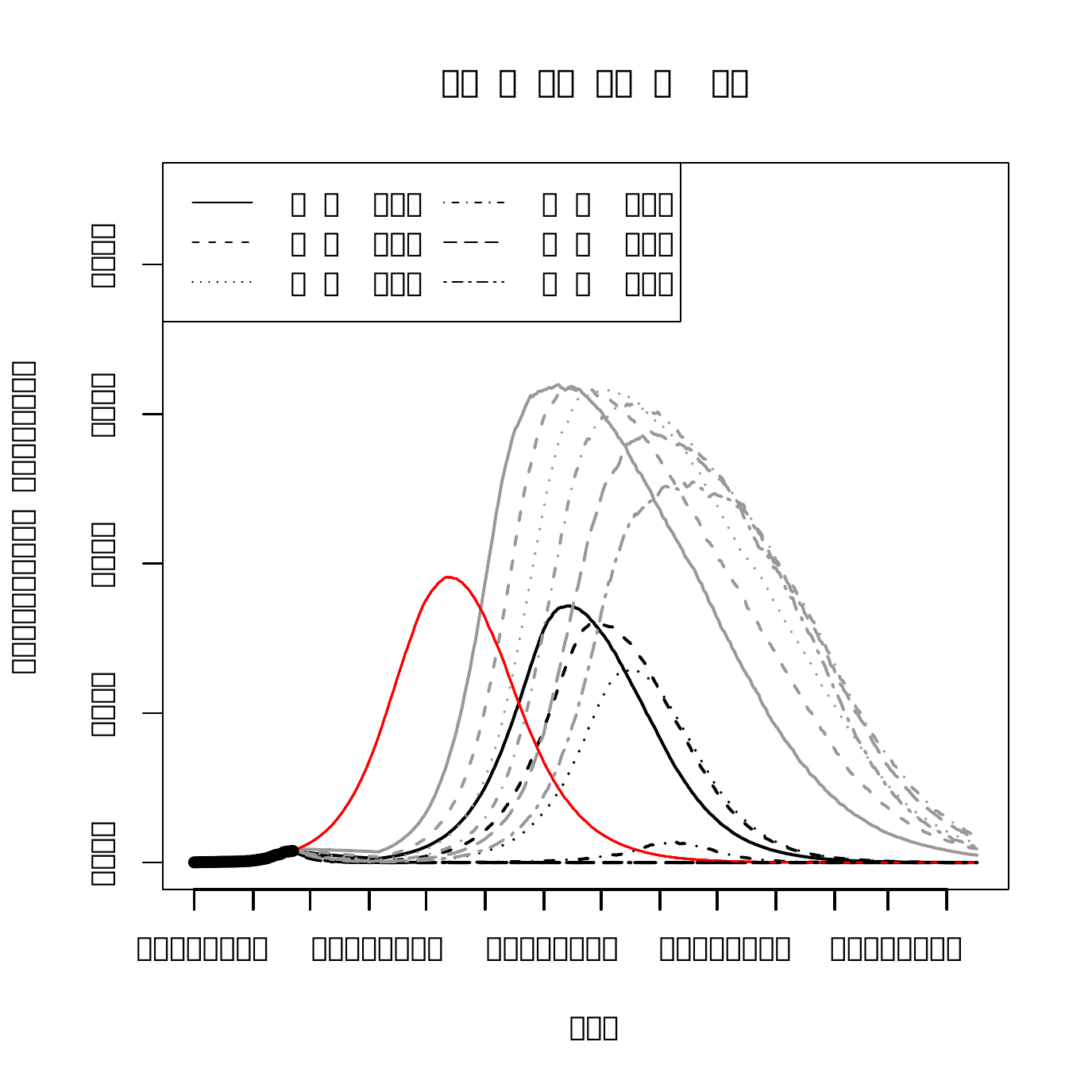}\\
\includegraphics[width=4.6cm]{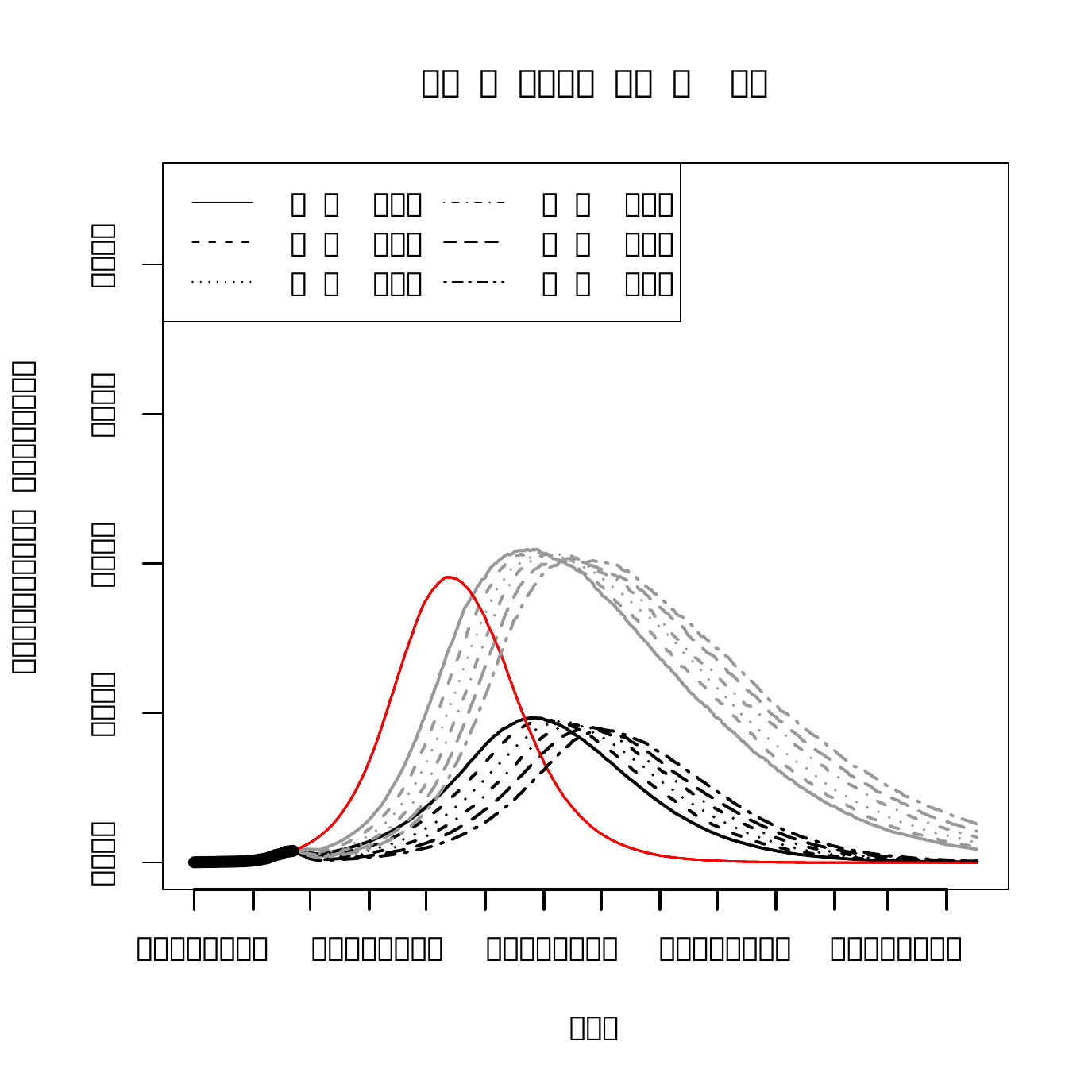}
\includegraphics[width=4.6cm]{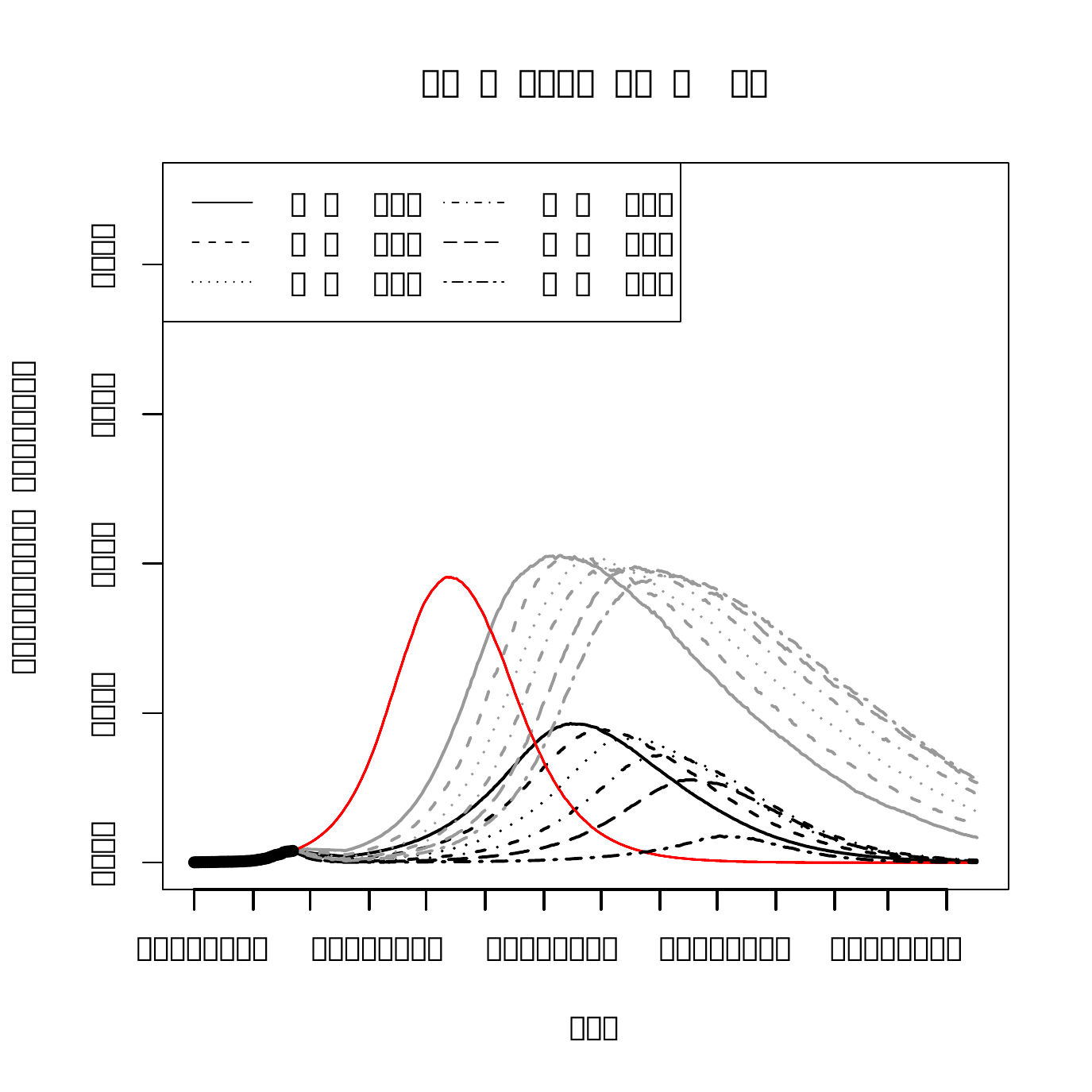}
\includegraphics[width=4.6cm]{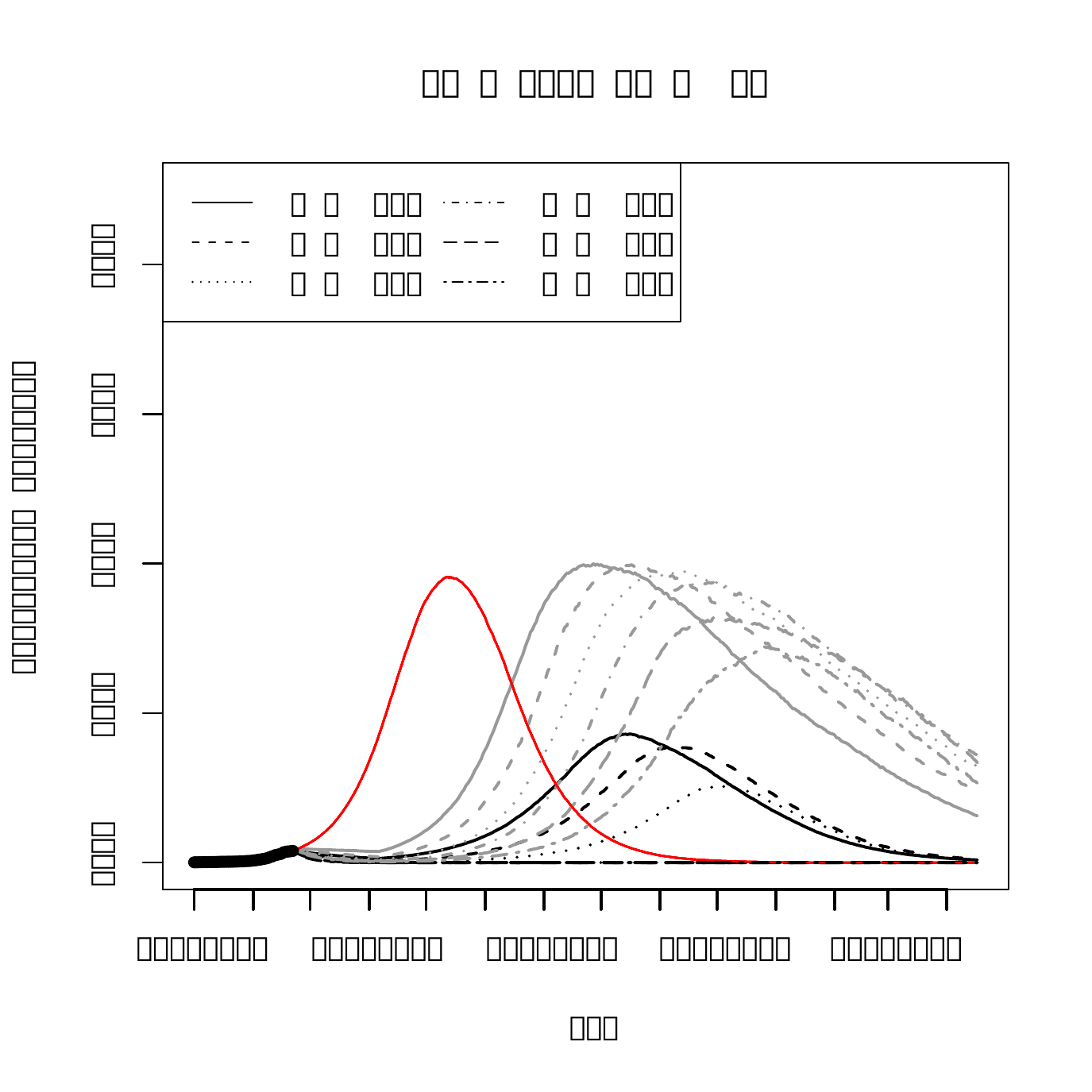}\\
\includegraphics[width=4.6cm]{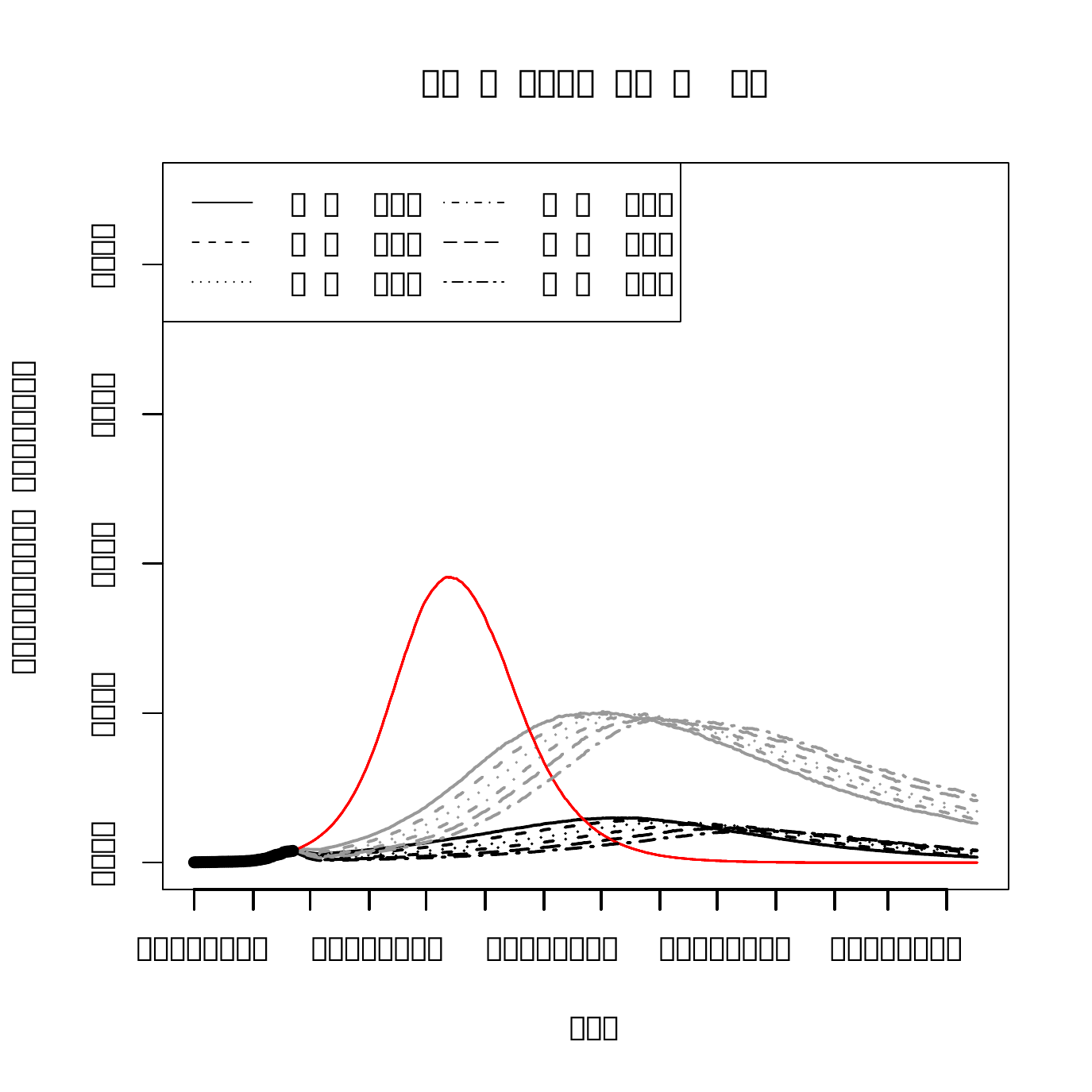}
\includegraphics[width=4.6cm]{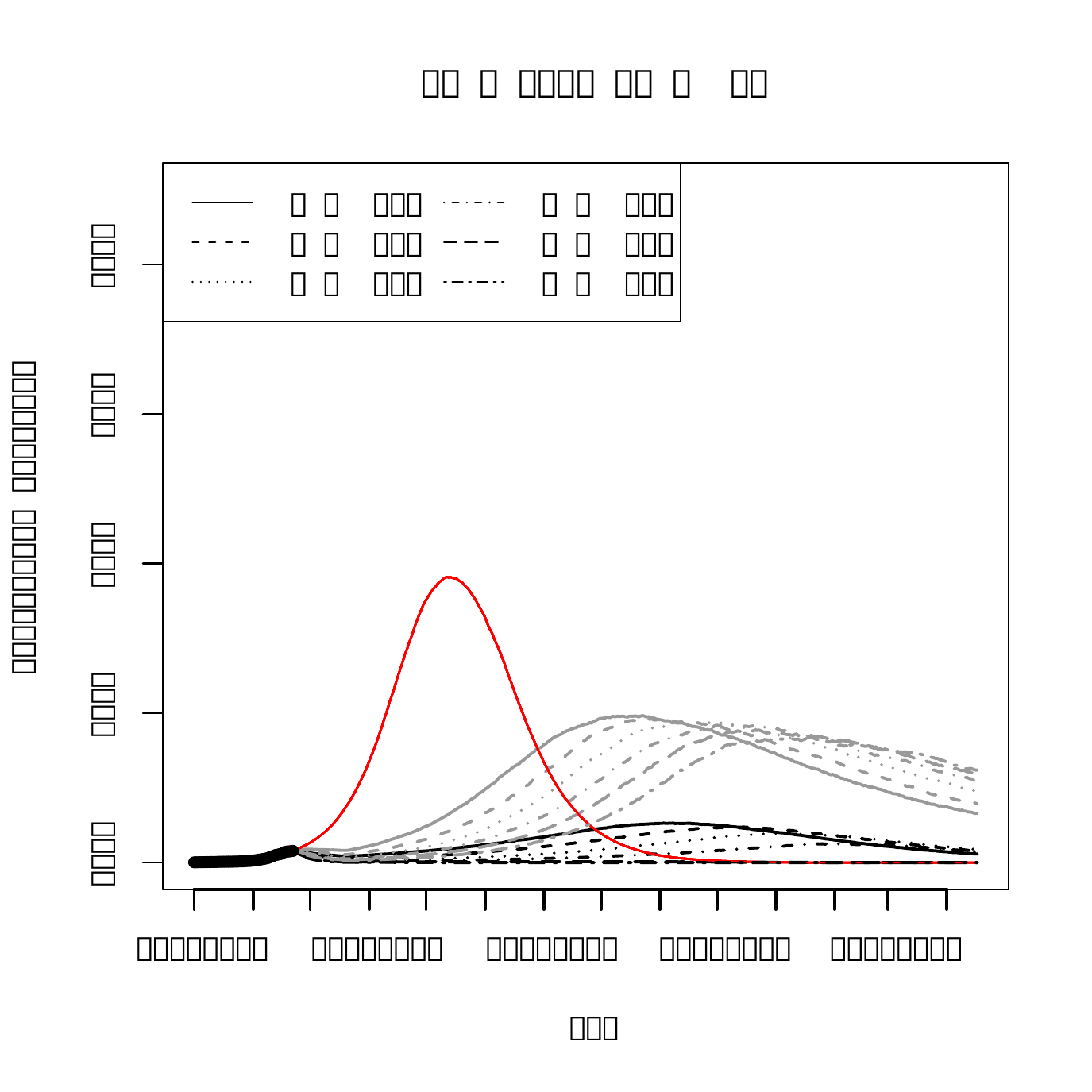}
\includegraphics[width=4.6cm]{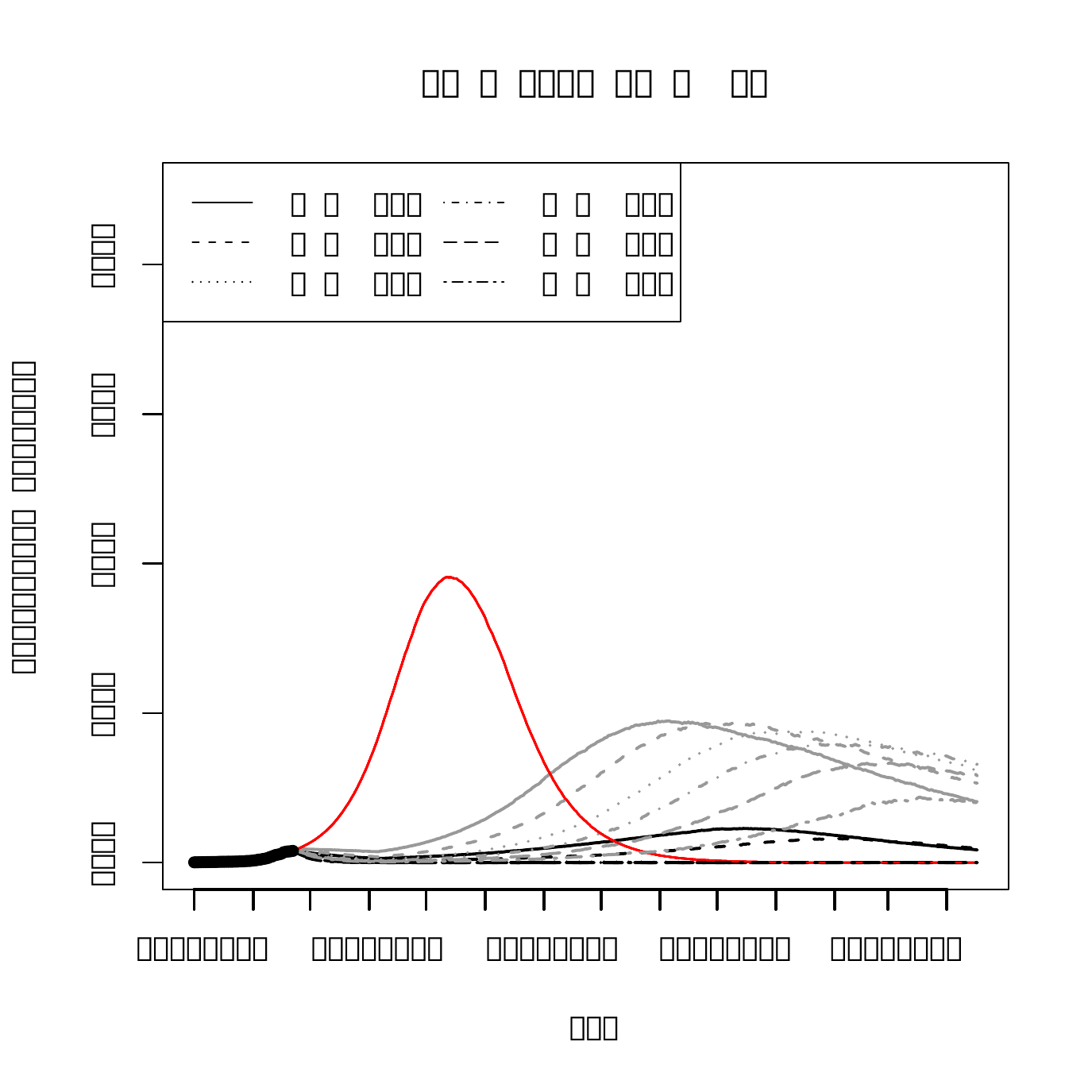}\\
\caption{Future prediction under the nine combinations $T^{\ast}$ (the period of the intervention) and $c^{\ast}$ (the multiplier for $\beta$ after the intervention) for $p=0.05$. 
The upper, middle and lower panels correspond to $c^{\ast}=1, 0.9$ and $0.8$, respectively. 
The red, black and grey curves respectively represent the future point prediction without intervention shown in Figure 2 of the main text, point prediction under each scenario and one-sided upper $95\%$ prediction intervals.  
 \label{fig:add-sim2}}
\end{figure}

\begin{figure}[htb!]
\centering
\includegraphics[width=4.6cm]{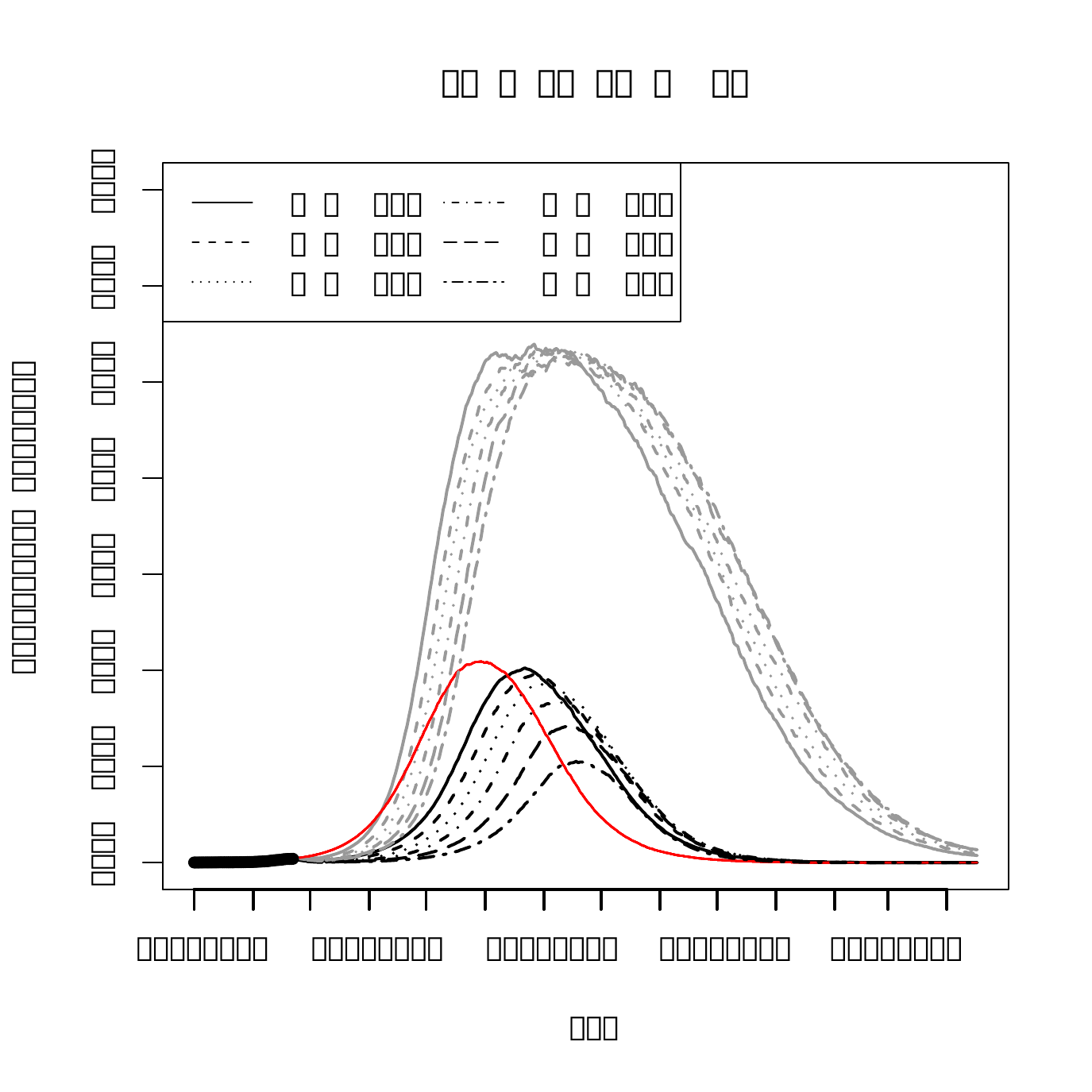}
\includegraphics[width=4.6cm]{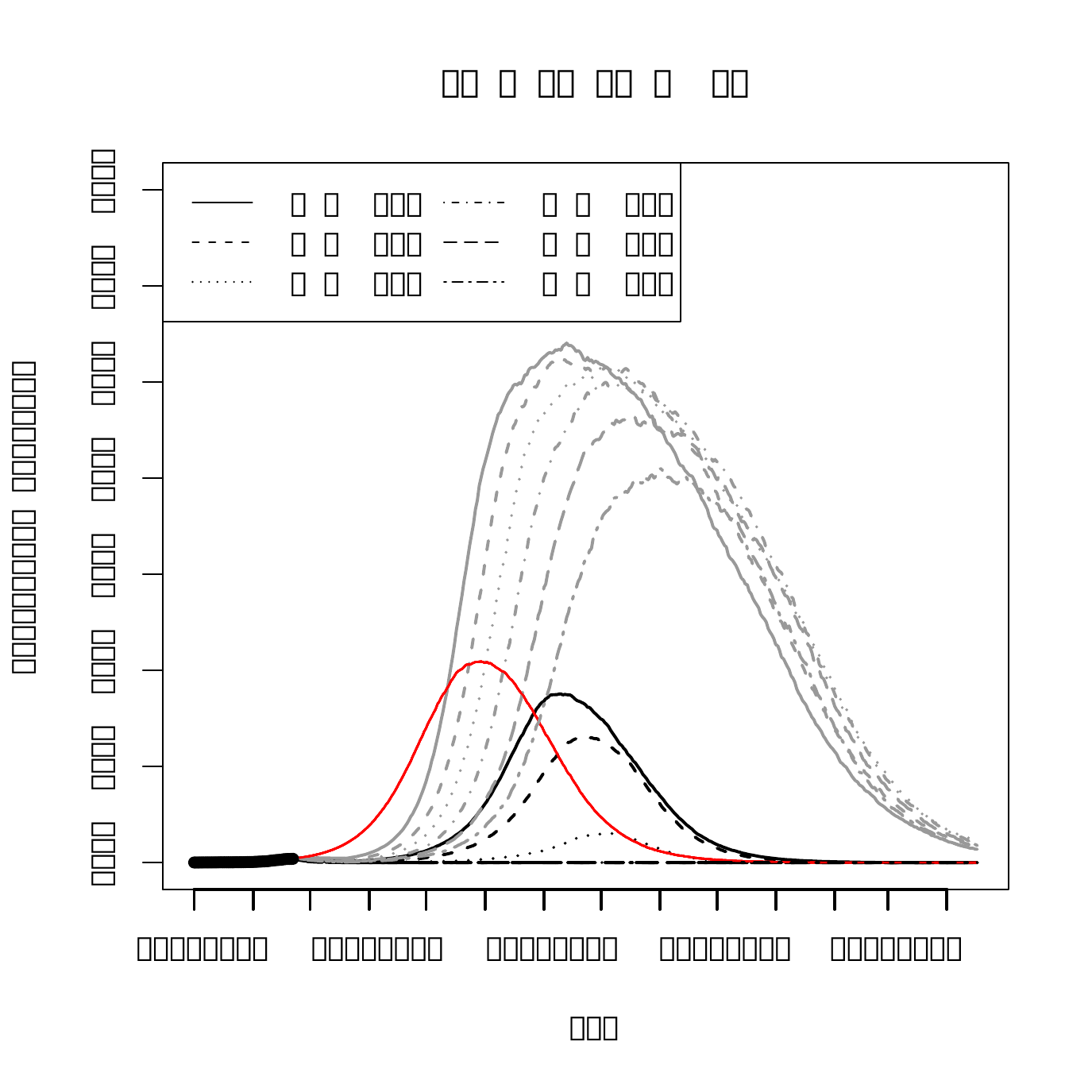}
\includegraphics[width=4.6cm]{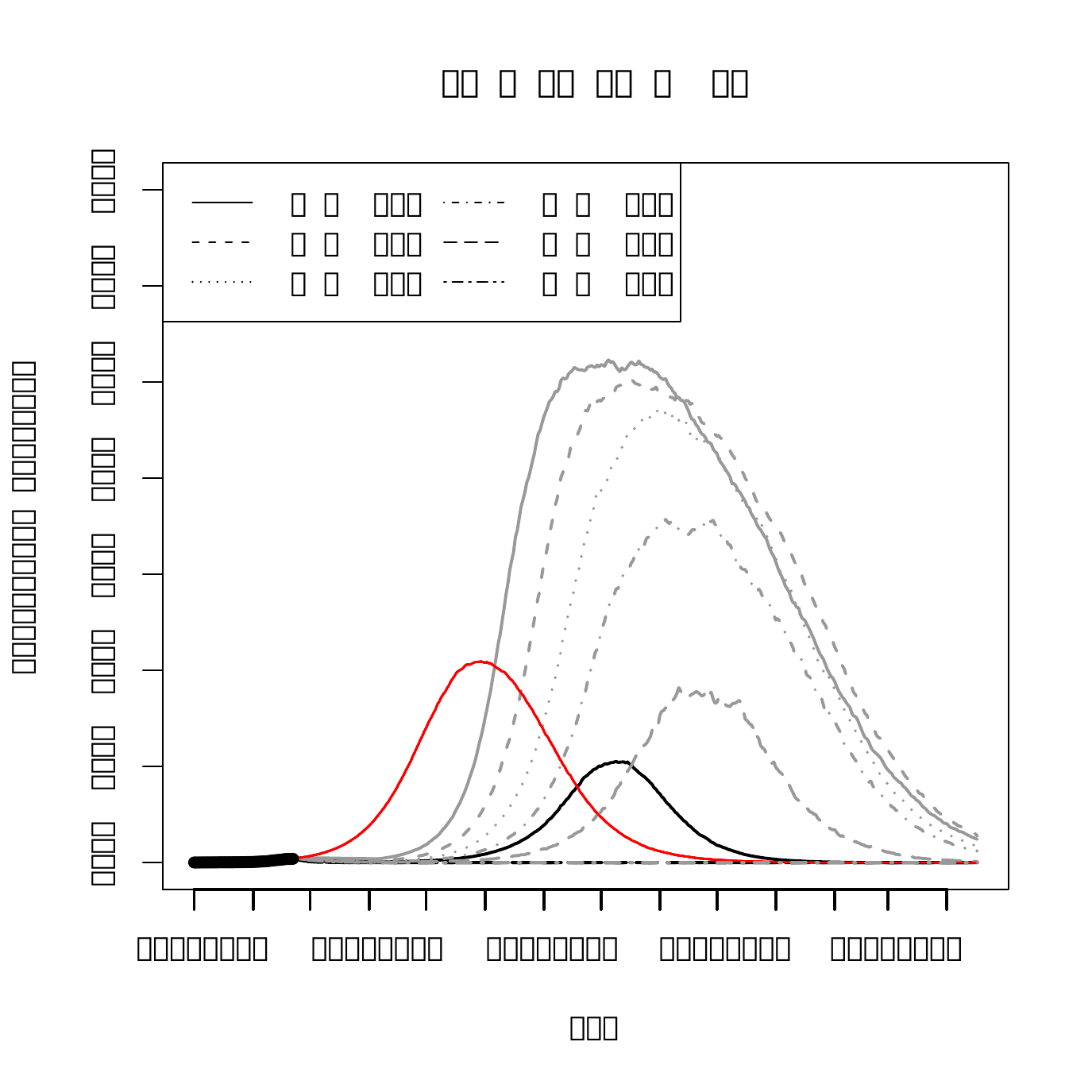}\\
\includegraphics[width=4.6cm]{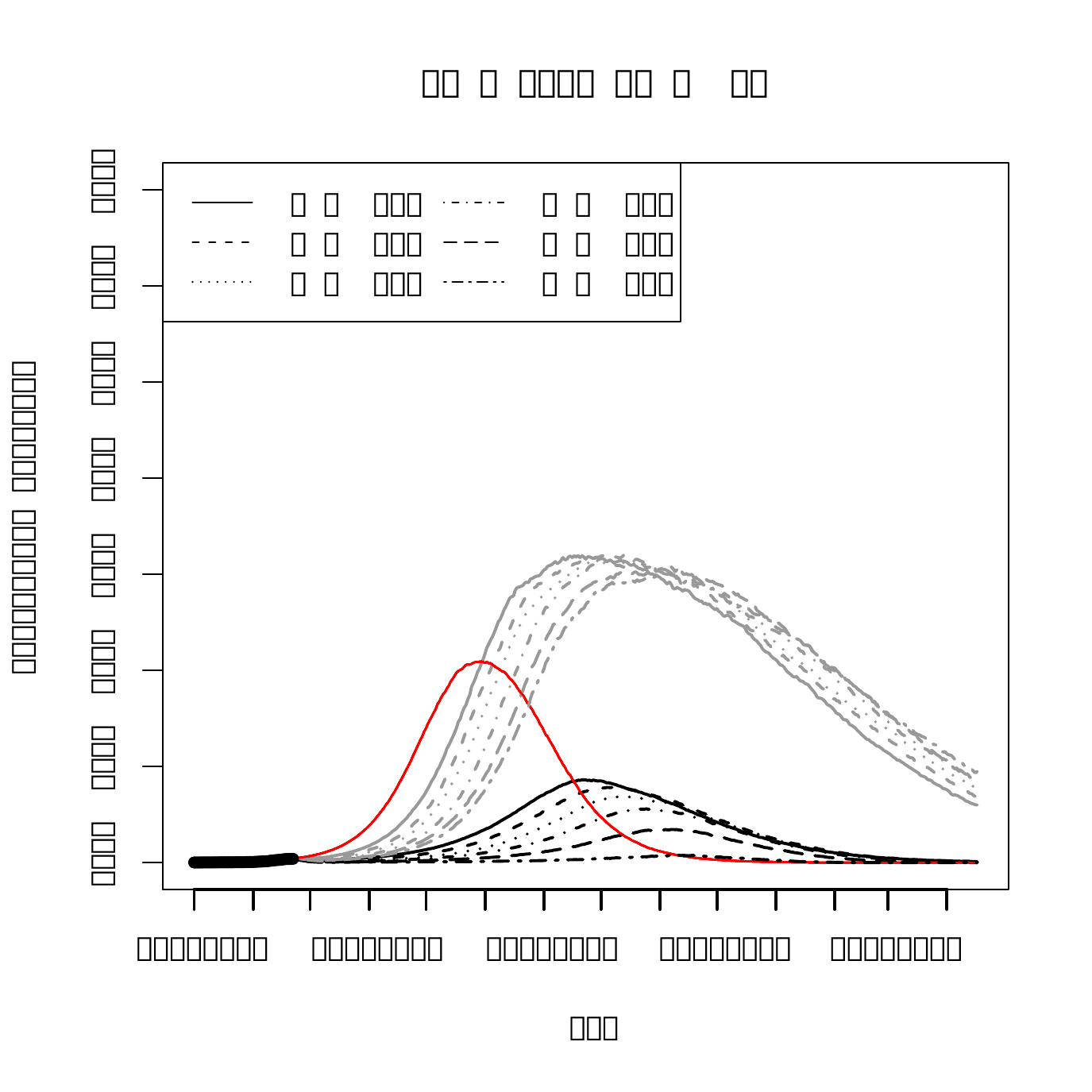}
\includegraphics[width=4.6cm]{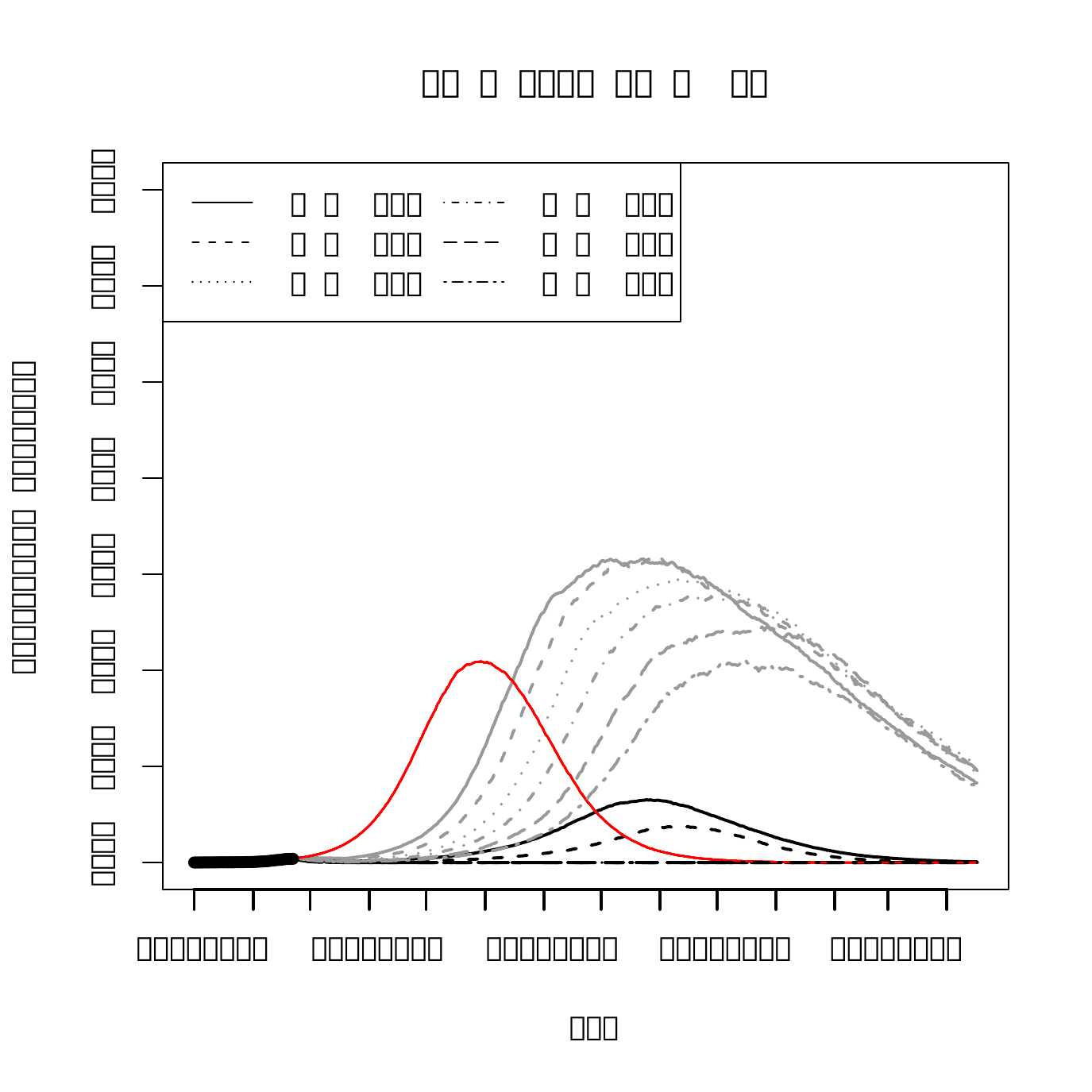}
\includegraphics[width=4.6cm]{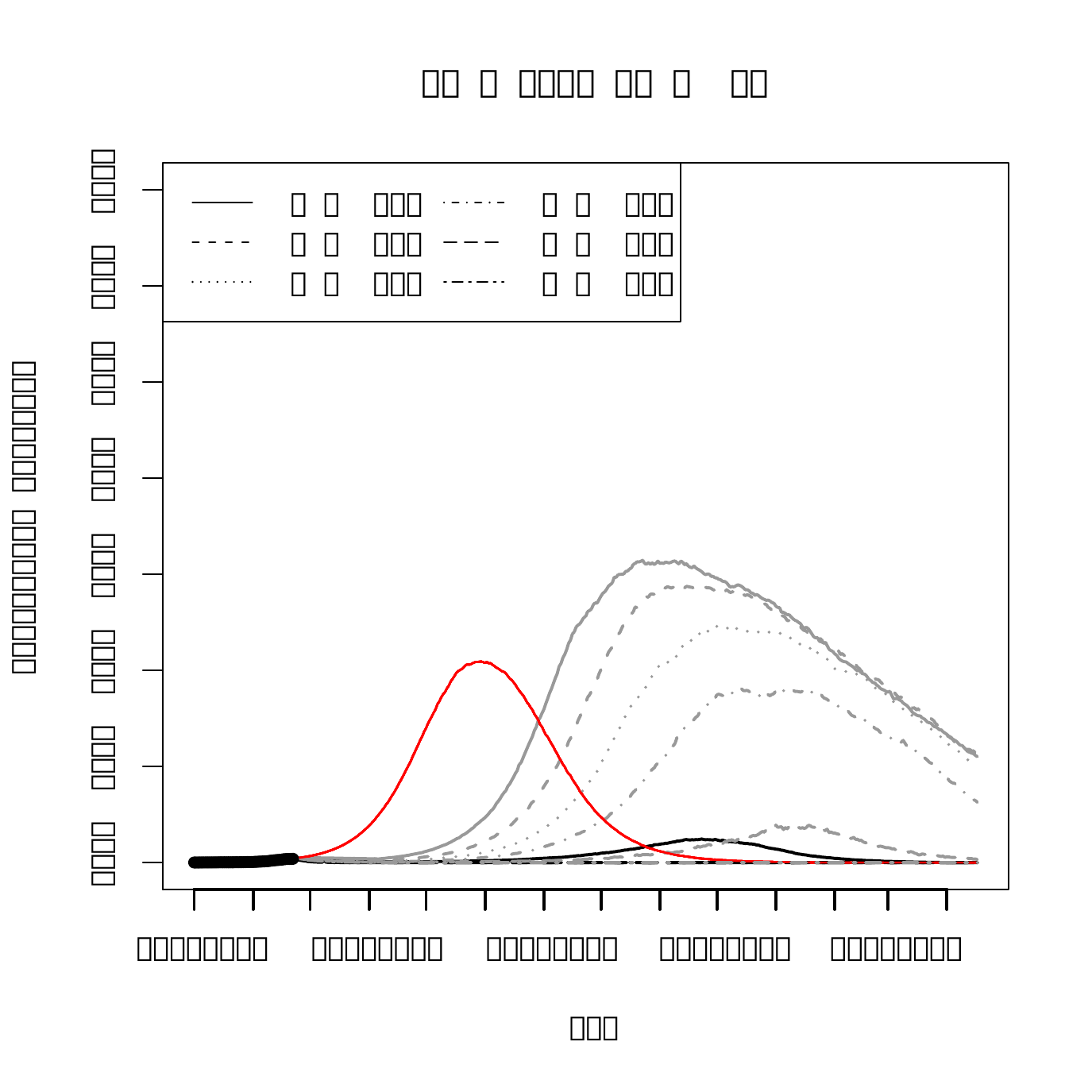}\\
\includegraphics[width=4.6cm]{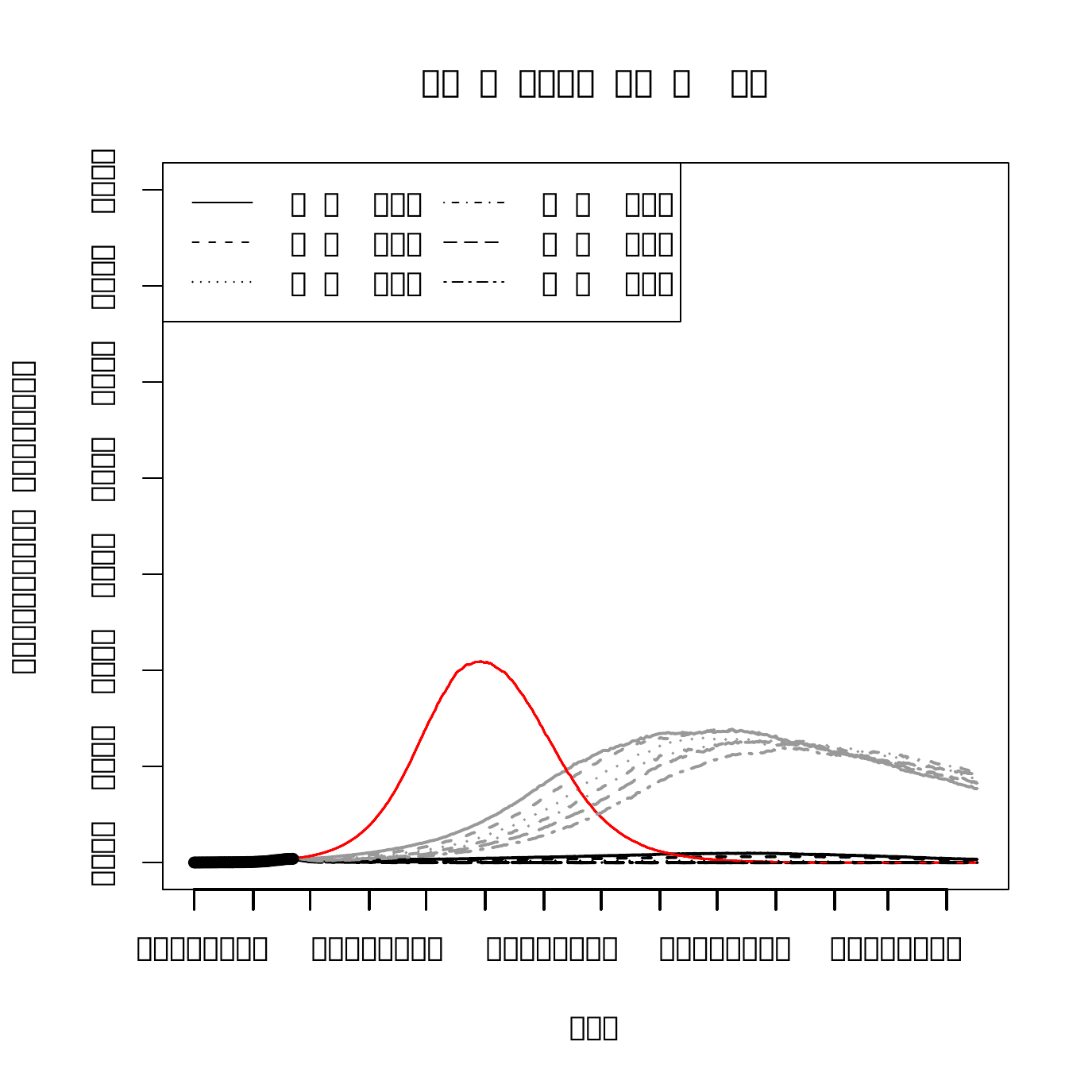}
\includegraphics[width=4.6cm]{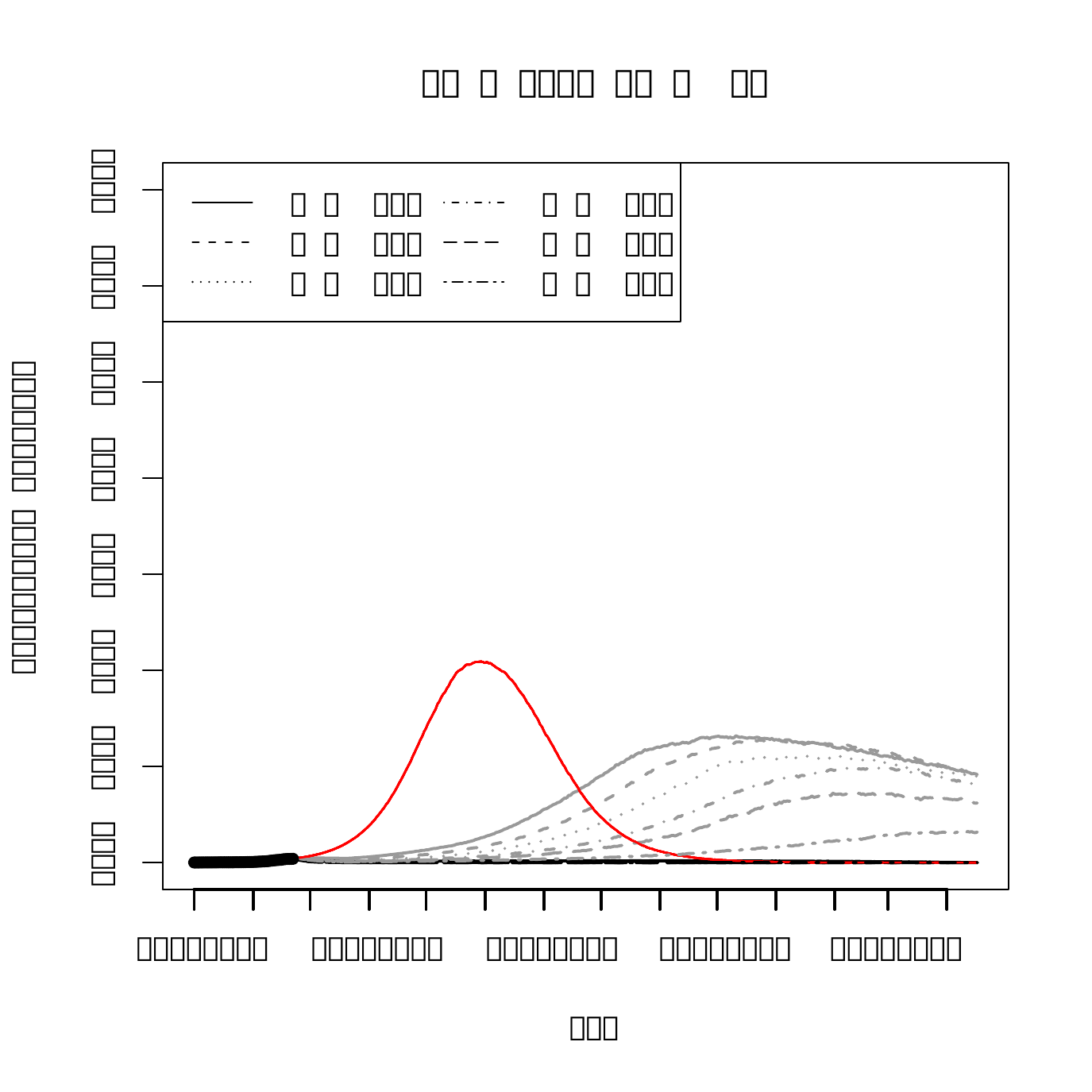}
\includegraphics[width=4.6cm]{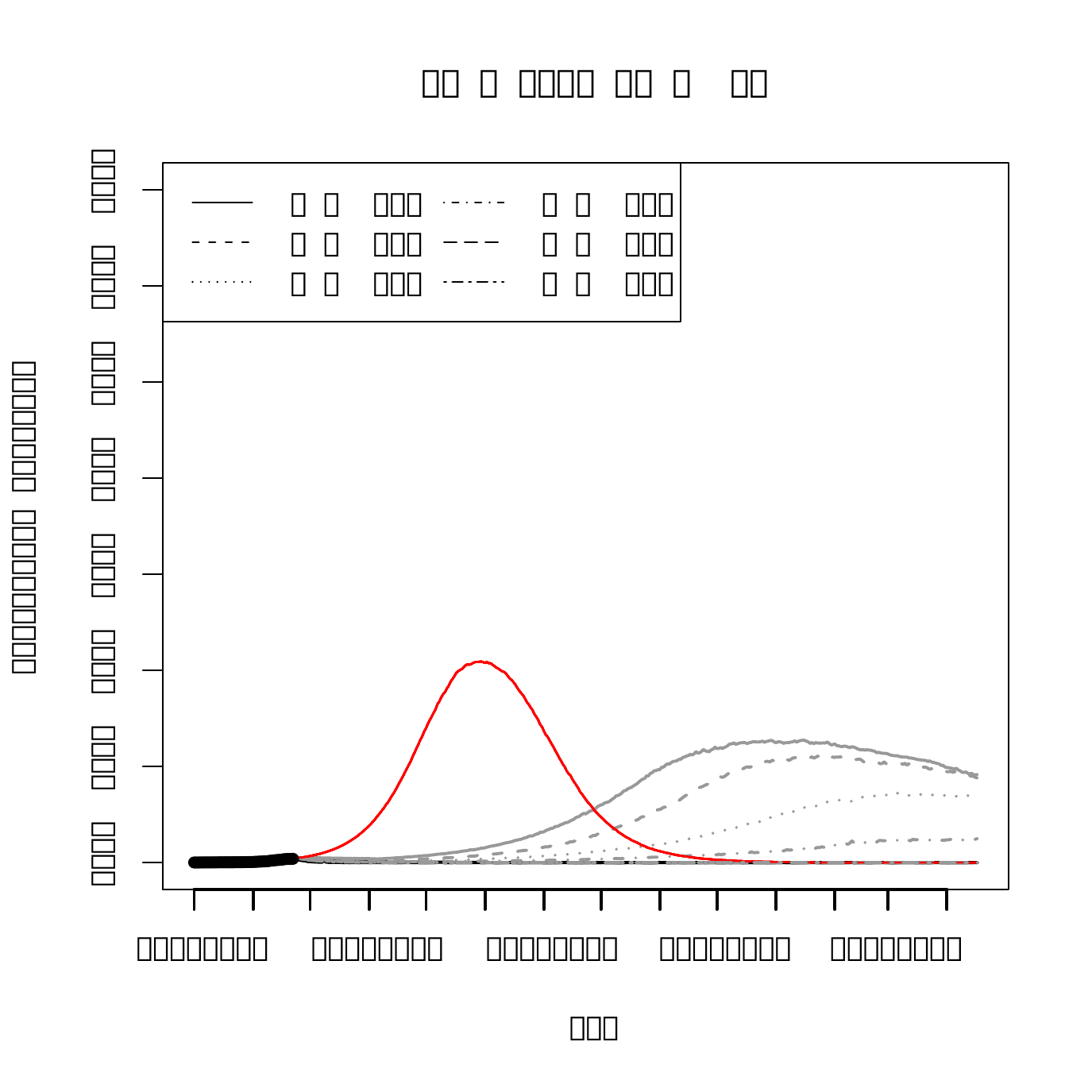}\\
\caption{Future prediction under the nine combinations $T^{\ast}$ (the period of the intervention) and $c^{\ast}$ (the multiplier for $\beta$ after the 
intervention) for $p=0.2$. 
The upper, middle and lower panels correspond to $c^{\ast}=1, 0.9$ and $0.8$, respectively. 
The red, black and grey curves respectively represent the future point prediction without intervention shown in Figure 2 of the main text, point prediction under each scenario and one-sided upper $95\%$ prediction intervals.  
 \label{fig:add-sim3}}
\end{figure}

Figure~\ref{fig:add-sim2} and \ref{fig:add-sim3} present the prediction results for $p=0.05$ and $0.2$ under the same scenarios in the main text for $p=0.1$. 
While the predicted timings of the peak and peak intensities vary depending on the choice of $p$, the figures provide the same epidemiological insights as discussed in Section 4.

\end{document}